\documentclass[notitlepage,nofootinbib,showpacs,unsortedaddress,aps,prd,12pt]{revtex4-1}
\usepackage[utf8x]{inputenc}
\usepackage{amsmath, graphicx, amsfonts, slashed}
\usepackage[normalem]{ulem}
\usepackage{color}
\usepackage{hyperref}

\numberwithin{equation}{section}

\newcommand{\rem}[1]{}
\newcommand{\ins}[1]{}

\newcommand{\cp}[3]{\ensuremath{\left( #1\! \times\! #2 \right)^{#3} }}
\newcommand{\vev}[1]{\ensuremath{\left< #1 \right> }}
\newcommand{\F}{\ensuremath{\mathcal{FT}}}

\newcommand{\var}[2][]{\ensuremath{\frac{\delta #1}{\delta #2}}}
\newcommand{\half}{{\textstyle{\frac 1 2}}}

\newcommand{\cpt}[3]{\ensuremath{\left( #1 \,{\widetilde\times}\, #2 \right)^{#3} }}
\newcommand{\Tr}{\ensuremath{{\rm Tr}}}
\newcommand{\drop}[1]{}

\newcommand{\eps}{\varepsilon}

\newcommand{\vph}{\varphi}

\newcommand{\beq}{\be}
\newcommand{\eeq}{\ee}
\newcommand{\p}{\partial}
\newcommand{\beqa}{\begin{eqnarray}}
\newcommand{\eeqa}{\end{eqnarray}}

\newcommand{\vac}{\ensuremath{|\text{vac}\rangle}}
\newcommand{\vevv}[1]{{\ensuremath{\langle\text{vac}|}#1{|\text{vac}\rangle}}}
\newcommand{\ket}[1]{\ensuremath{\vert #1 \rangle}}

\newcommand{\braket}[2]{\ensuremath{\left< #1 \middle\vert #2 \right> }}

\newcommand{\Wphys}{{\cal W}_{\rm phys}}

\newcommand{\be}{\begin{equation}}
\newcommand{\ee}{\end{equation}}
\renewcommand{\eqref}[1]{Eq.~(\ref{#1})}

\renewcommand{\L}{\ensuremath{\mathcal L}}

\begin{document}

\title{BRST cohomology and physical space of the GZ model}

\author{Martin Schaden}
\email{mschaden@rutgers.edu}
\affiliation{Department of Physics, Rutgers, The State University of New Jersey, 101 Warren Street, Newark, New Jersey - 07102, USA }

\author{Daniel Zwanziger}
\email{dz2@nyu.edu}
\affiliation{Physics Department, New York University, 4 Washington Place, New York, NY 10003, USA}

\begin{abstract}
\noindent{\bf Abstract:} We address the issue of Becchi-Rouet-Stora-Tyutin (BRST) symmetry breaking in the Gribov-Zwanziger (GZ) model, a local, renormalizable, non-perturbative approach to QCD.  Explicit calculation of several examples reveals that BRST symmetry breaking apparently afflicts the unphysical sector of the theory, but may be unbroken where needed, in cases of physical interest.  Specifically, the BRST-exact part of the conserved energy-momentum tensor and the BRST-exact term in the Kugo-Ojima confinement condition both have vanishing expectation value.  We analyze the origin of the breaking of BRST symmetry in the GZ model, and obtain a useful sufficient condition that determines which operators preserve BRST.  Observables of the GZ theory are required to be invariant under a certain group of symmetries that includes not only BRST but also others.  The definition of observables is thereby sharpened, and excludes all operators known to us that break BRST invariance.  We take as a hypothesis that BRST symmetry is unbroken by this class of observables.  If the hypothesis holds, BRST breaking is relegated to the unphysical sector of the GZ theory, and its physical states are obtained by the usual cohomological BRST construction.  The fact that the horizon condition and the Kugo-Ojima confinement criterion coincide assures that color is confined in the GZ theory.
\end{abstract}
\maketitle

\section{Introduction}
\label{introduction}

The GZ model is a non-perturbative approach to QCD that provides a cut-off at the Gribov horizon~\cite{Gribov:1977wm} by means of a local, renormalizable, continuum action~\cite{Zwanziger:1989mf, Zwanziger:1993}.  For this reason the gap equation that determines the value of a parameter $\gamma$, the Gribov mass, is known as the ``horizon condition".   For a review, see~\cite{Sobreira:2004, Vandersickel:2012}. 

It is a remarkable fact that the horizon condition and the famous Kugo-Ojima confinement criterion~\cite{Kugo-Ojima, Kugo:1995km} are the identical statement,
\beq
\label{horizonconditiona}
- i \int d^dx\vev{(D_\mu c)^a(x) (D_\mu \bar c)^a(0) } = d (N^2 -1),
\eeq
where $c^d$ and $\bar c^d$ are the Faddeev-Popov ghosts, $(D_\mu)^{ad} = \p_\mu \delta^{ad} + g f^{abd} A_\mu^b$ is the gauge-covariant derivative in the adjoint representation of $SU(N)$, and $A_\mu^b$ is the gluon field in Landau gauge.  This is promising for the confinement problem, because the Kugo-Ojima criterion is a sufficient condition for color confinement, and the horizon condition assures that this condition is satisfied in the GZ approach.  Although the identity of these two conditions has been noted for some time~\cite{Dudal:0904,Mader:2014}, its consequences have remained obscure because the Kugo-Ojima confinement criterion requires BRST symmetry~\cite{BRST}  to be unbroken, whereas the GZ vacuum breaks BRST symmetry.

This breaking is manifested by the non-zero vacuum expectation value of a BRST-exact quantity such as
\be
\left\langle {\rm vac}|  \{ Q_B,  \bar\omega \} |{\rm vac} \right\rangle \neq 0, 
\eeq   
where $\bar\omega$ is an auxiliary ghost field of GZ theory, $Q_B$ is the BRST charge, and $\vac$ is the vacuum state.  It follows formally that $Q_B \vac \neq 0$. The possibility that BRST may be dynamically broken due to the Gribov ambiguity was first considered by Fujikawa~\cite{Fujikawa:1982ss} and is discussed in~\cite{Schaden:1996}.  A gauge theory with dynamically broken BRST is not standard.  In perturbative Faddeev-Popov theory, physical states $\ket{\rm phys}$ are precisely characterized by the condition $Q_B \ket{\rm phys} = 0$ and so, according to the standard paradigm, the vacuum of GZ theory would not be a physical state. Clearly a different construction is required if the GZ theory is to be consistent.  

For a hint on how to proceed, consider the Ward-Takahashi identity that expresses conservation of the energy-momentum tensor,
\beq
\langle \ \p_\mu T_{\mu \nu} \  I(A) \ \rangle = \left\langle {\delta I \over \delta A_\mu^b} F_{\mu \nu}^b \right\rangle,
\eeq
where $I(A)$ is any gauge-invariant functional of the gauge connection $A$, and the energy-momentum tensor is given by,
\beq
\label{TplusXi}
T_{\mu \nu} = T_{\mu \nu}^{\rm YM} + s \Xi_{\mu \nu}.
\eeq
Here
\beq
\label{TYM}
T_{\mu \nu}^{\rm YM} \equiv  F_{\mu \lambda}^a F_{\nu \lambda}^a - { 1 \over 4} \delta_{\mu \nu} F_{\kappa \lambda}^a F_{\kappa \lambda}^a\ ,
\eeq
is the Yang-Mills energy-momentum tensor of Maxwellian form, and $F_{\mu \nu}^a = \p_\mu A_\nu^a - \p_\nu A_\mu^a + g f^{abc}A_\mu^b A_\nu^c$.  This Ward identity holds, with different $\Xi_{\mu \nu}$, in both the Faddeev-Popov and the GZ theories, because it is a consequence of the translation invariance of the action.  In Faddeev-Popov theory, the $s$-exact contribution to the Ward identity vanishes, $\vev{s \Xi_{\mu \nu} \  I(A)} = \vev{ s[ \Xi_{\mu \nu} \  I(A) ] } = 0$, to every order in perturbation theory and the Ward identity reads
\beq
\label{Wardidentity}
\langle \ \p_\mu T_{\mu \nu}^{\rm YM} \  I(A) \ \rangle = \left\langle {\delta I \over \delta A_\mu^b} F_{\mu \nu}^b \right\rangle .
\eeq
It involves only gauge-invariant quantities and holds in every gauge\footnote{This Ward identity presumably also holds in the continuum limit of lattice gauge theory~\cite{Caracciolo:1990}, but this is difficult to show because the translation-group of the lattice is discrete, whereas the Ward identity is a consequence of Noether's theorem for continuous (Lie) groups.}.  We are loth to give up this physical identity that relies on the vanishing of the vacuum expectation value of an $s$-exact quantity, $ \langle s [ \p_\mu \Xi_{\mu \nu} \  I(A) ] \rangle = 0$, which is assured when BRST is unbroken.  Some BRST-exact operators, $sF$, do have vanishing expectation value in GZ theory, $\vev{sF}=0$, but it can be difficult to ascertain whether or not it does for a given $sF$.  That generally depends on whether $\gamma$ has the precise value fixed by the (non-perturbative) horizon condition, \eqref{horizonconditiona}.  However, we have verified by direct calculation that the BRST-exact part of the energy-momentum tensor has vanishing expectation value $\langle s\Xi_{\mu \nu} \rangle = 0$ when the horizon condition holds.  We also found that, to leading order, the BRST-exact term [see \eqref{quartet}  below] in the derivation of the Kugo-Ojima criterion~\cite{Kugo-Ojima} has a vanishing expectation value.  These results suggest that in the GZ model BRST symmetry may be preserved precisely where it is needed, although it fails for some unphysical expectation-values such as $\vev{s \bar\omega} \neq 0$.

The GZ theory has auxiliary ghosts and a rich set of unphysical symmetry generators $Q_X$ that do not appear in Faddeev-Popov theory.  Requiring that all physical observables be invariant under these symmetries, $[Q_X, F] = 0$, in addition to the BRST symmetry $[Q_B, F] = 0$, sharpens the definition of an observable\footnote{In Faddeev-Popov theory, physical observables $F$ in fact are required to also commute with ghost number $[Q_{\cal N}, F] = 0$.}.  We propose the hypothesis that in the GZ theory BRST symmetry remains unbroken, $\vev{s F} = 0$, for all $s$-exact observables $sF$.  This relegates the breaking of BRST symmetry to the unphysical sector of the GZ-theory and is sufficient for the familiar BRST construction of physical states as the cohomology of the BRST operator.

 Let us briefly address some issues that have been raised about the GZ action.  It was originally derived~\cite{Zwanziger:1989mf, Zwanziger:1993} to provide a cut-off at the Gribov horizon. This procedure has been criticized because there are Gribov copies within the Gribov horizon.  However, the proposed local action has interesting properties, such as renormalizability and renormalizability of the horizon condition and the coincidence of the horizon condition with the Kugo-Ojima confinement criterion, which make it worthy of study even if the model should turn out to be approximate.  Subsequently, the same local action was rederived by an entirely different line of reasoning~\cite{Schaden:1994}.  One starts in the conventional way with an $s$-exact extension of the Yang-Mills action.  A redefinition of the fields, the Maggiore-Schaden (MS) shift, then produces the GZ action, and the horizon condition arises as a gap equation for the new vacuum.  In this approach, BRST symmetry is spontaneously broken by the new vacuum, instead of being explicitly, though softly, broken by the GZ action. 

The distinction arises from two different definitions of the BRST symmetry.  In the present article we are concerned with a BRST symmetry that is an exact, but spontaneously broken, symmetry of the GZ action.  The alternative BRST symmetry is explicitly, though softly, broken by the GZ action~\cite{Sobreira:2004}.   Explicit soft BRST symmetry breaking has recently been proposed~\cite{Capri:2014} as a mechanism that phenomenologically describes the confinement of matter. The breaking of BRST symmetry was recently  studied numerically~\cite{Cucchieri:2014}.  An approach to the restoration of BRST symmetry is presented in \cite{Dudal:2010hj}, following ideas in \cite{Sorella:2009vt} and \cite{Kondo:2009qz}.  The explicit soft breaking of BRST symmetry might not be consistent with Batalin-Vilkovisky quantization~\cite{Lavrov:2011} (for more recent results see \cite{Lavrov:2013boa,Reshetnyak:2013bga}).  Spontaneous breakin
g of BRST symmetry has been questioned~\cite{Dudal:2012sb} on the ground that it apparently goes beyond standard quantum field theory.  The issue here is that it should be mathematically well defined.  This point is addressed in Sect.~\ref{surface equation} of the present work, where the GZ action is quantized in a finite, periodic box (see~\cite{Vandersickel:2012}, p. 226).  The analysis at finite volume yields a criterion for which operators $sF$  preserve BRST symmetry $\vev{sF} = 0$ in the infinite-volume limit.

Perturbative calculations up to two loops of the GZ action in three~\cite{Gracey:2010df} and four~\cite{Ford:2009ar,Gracey:2009zz,Gracey:2009mj} Euclidean dimensions as well as a non-perturbative infrared analysis~\cite{Huber:2009tx} show that the gluon propagator of this theory vanishes at long wavelengths. The propagators of the Faddeev-Popov (FP)  and of auxiliary fermi ghost  obtained by solving the Dyson-Schwinger-Equations (DSE) are identical,\footnote{This is a consequence of the symmetry generated by $Q_R$ of \eqref{QR}.}  and have an  enhanced singularity at vanishing momentum. In the GZ-theory this is the only solution to the DSE \cite{Huber:2010} so far, and the enhancement is due to the horizon condition.  Ghost- and gluon- propagators with the same infrared exponents were also found in the numerical solution to the DSE of the FP-theory\cite{Fischer:2008uz}.  This solution to the DSE supports the Kugo-Ojima confinement scenario.  It is consistent with lattice simulations in Landau gauge in two~\cite{Maas:2007uv, Cucchieri:2007rg, Cucchieri:2010xr}, but not in three and four~\cite{Cucchieri:2007rg,Cucchieri:2010xr,Sternbeck:2007,Bogolubsky:2007ud,Bogolubsky:2009dc,Bornyakov:2009ug,
Spielmann:2011} dimensions\footnote{The Gribov scenario is consistent with numerical calculations in Coulomb gauge in 4d~\cite{Cucchieri:2002, Langfeld:2004, GOZ:2005}.  The considerations concerning the Landau gauge that are reported in the present article are expected to carry over to the GZ action in Coulomb gauge~\cite{Zwanziger:2007bz, Zwanziger:2006prz}.  The calculation in Appendix \ref{testBRST}, shows that the $s$-exact part of the energy-momentum tensor $T_{\mu \nu} = T_{\mu \nu}^{\rm YM} + s \Xi_{\mu \nu}$ has vanishing expectation-value $\vev{s \Xi_{\mu \nu}} = 0$. This also holds in Coulomb gauge.}.  We do not offer a resolution of this matter in the present article, but note that the value of the ghost dressing function at vanishing momentum is a gauge-dependent quantity~\cite{Fischer:2008uz,Dudal:2014rxa}.  It parametrizes different gauges within the family of Landau gauges. Lattice evidence for the dependence of Landau gauge propagators on additional constraints was obtained in~\cite{Maas:2008, Sternbeck:2012mf}.   This observation perhaps helps to resolve the discrepancy between the far infrared behavior of the gluon and ghost propagators in Landau gauge of the lattice and of GZ theory for space-time dimensions $d >2$. 

The present article is organized as follows.  For completeness and because it is not well known, the MS shift is used to derive the GZ action in Sect.~\ref{MSshift},  and the horizon condition for the new vacuum is obtained in Sect.~\ref{newvacuum}.  Sect.~\ref{BRSTbreaking} is devoted to the analysis of BRST breaking:  BRST breaking is exhibited in Sect.~\ref{LostBRST};  the GZ action is quantized in a periodic box and the BRST-breaking term is expressed as an integral over the surface of the box in Sect.~\ref{surface equation};  a sufficient condition for an operator to preserve BRST symmetry is derived in Sect.~\ref{sufficientcondition}.  The physical state space of the GZ theory is constructed in Sect.~\ref{BRST-GZconstruction}:  observables are identified as those functionals that commute with all phantom symmetries in Sect.~\ref{physoperators}, and in Sect.~\ref{hypothesis} we introduce the hypothesis that BRST symmetry is not broken by $s$-exact observables; in Sect.~\ref{cohomology} the physical Hilbert space of the model is reconstructed from its observables and identified with the cohomology of the BRST operator; in Sect.~\ref{positivity} the positivity of the Euclidean inner product of physical states is established. We derive the energy-momentum tensor of the theory in Sect.~\ref{EMtensor}, and in Sect.~\ref{testhypothesis}, we prove that  the expectation value of the $s$-exact part of the energy-momentum tensor vanishes.  We capitalize on this by computing the trace anomaly of the GZ-theory to one loop in Sect.~\ref{traceanomaly}. The anomaly at one loop has a finite negative value and establishes that the vacuum with $\gamma>0$ has lower energy-density.  In Sect.~\ref{KO-BRST} we find that the $s$-exact term in the derivation of the Kugo-Ojima equation has vanishing vacuum expectation value.  Sect.~\ref{conclusion} gives our summary and  conclusions.  The unphysical symmetries are compiled in Appendix~\ref{phantomsymm}. A special case of the surface equation of Sect.~\ref{surface equation} is considered in Appendix~\ref{checksurfaceeq}.  The criterion of Sect.~\ref{sufficientcondition} is applied in Appendix \ref{applycriterion} to a simple operator that preserves BRST symmetry when the horizon condition holds.    An alternative criterion to  the surface equation is derived in Appendix~\ref{alternativemeansPi}.    In Appendix~\ref{testBRST} we give a second proof by direct evaluation that $\vev{T_{\mu \nu}} = \vev{T_{\mu \nu}^{\rm YM}}$.

\section{Local action by the MS shift}\label{MSshift}

The Faddeev-Popov quantization of Yang-Mills theory in Landau gauge is defined by the Lagrangian density,
\begin{align}
 \L^{\rm FP} &=  \L^{\rm YM} + s \left( i\p_\mu \hat{\bar c}\cdot A_\mu\right)
 \nonumber\\
&= {1 \over 4} F_{\mu \nu}^2 + i \p_\mu \hat b\cdot A_\mu  - i \partial_\mu\hat{\bar c}\cdot D_\mu c \,,
\label{ko_lag}
	\end{align}
where $F_{\mu \nu} = \p_\mu A_\nu - \p_\nu A_\mu + A_\mu\times A_\nu$ is the Yang-Mills field strength. The connection $A^a_\mu$ as well as the Nakanishi-Lautrup and Faddev-Popov ghost fields $\hat b^a, c^a$ and $\hat{\bar c}^a$ are all fields in  the adjoint representation of the global  $SU(N)$ color group. Color components are represented by Latin superscripts. To streamline notation we adopt the convention that $X\cdot Y\equiv \sum_a X^a Y^a$ and $(X\times Y)^a\equiv \sum_{bc} g f^{abc} X^b Y^c$, where $f^{abc}$ are the $su(N)$ structure constants and $g$ is the gauge coupling. In this notation the gauge-covariant derivative in the adjoint representation is  $D_\mu X=\p_\mu X+ A_\mu\times X$.  

The nilpotent  BRST transformation is given by
\begin{align}\label{ko_BRST}
      s A_\mu & = D_\mu c \, , & sc &  = -\frac{1}{2} \cp{c}{c}{} \, , \nonumber \\ 
      s \hat{\bar c} & = \hat b\, , & s \hat b &= 0 \,, 
\end{align}
and is readily extended to covariantly coupled matter, with $s^2 = 0$.

A number of quartets of auxiliary ghosts, $(\phi_B, \bar\phi_B, \omega_B, \bar\omega_B)$ are introduced to localize the (otherwise non-local) cut-off at the Gribov horizon~\cite{Zwanziger:1989mf}. The index $B$ labels the quartets. $\phi_B$ and $\bar\phi_B$ are a bose ghost pair, and $\omega_B$ and $\bar\omega_B$  a corresponding pair of fermi ghosts. The auxiliary ghosts are in the adjoint color representation and  the BRST operator acts trivially on each quartet,
\begin{align}\label{sonaux}
s \phi_B &= \omega_B &  s \omega_B &= 0 \nonumber \\
s \bar\omega_B  &=  \bar\phi_B &   s \bar\phi_B &= 0.
\end{align}

In GZ theory the Yang-Mills Lagrangian density is similarly extended by an $s$-exact term, and takes the form,
\beq
\label{totallagrangian}
{\cal L} \equiv {\cal L}^{\rm YM} + {\cal L}^{\rm gf} =  {\cal L}^{\rm YM} + s \Psi
\eeq
where,
\beq
\label{PsiB}
{\cal L}^{\rm gf}=s\Psi, \ \text{with }\ \ \Psi \equiv  i \p_\mu \hat{\bar c}\cdot A_\mu + \p_\mu \bar\omega_B\cdot D_\mu \phi_B .
\eeq
Because ${\cal L}^{\rm gf}$ is $s$-exact, it should not change the physics. This is seen by formally integrating out the auxiliary ghosts: for each Faddeev-Popov determinant arising from integrating over a pair of fermi ghosts, one obtains a compensating inverse Faddeev-Popov determinant upon integration of a pair of bose ghosts.

The index set of the auxiliary ghosts is written as a pair $B = (\nu, b)$, where $b$ is an index that takes values in the adjoint representation of an $su(N)$ ``flavor" algebra  (not to be confused with physical flavor), and $\nu$ is interpreted as a vector index. Thus $\phi^a_B = \phi^a_{\nu b}$, and likewise for all the auxiliary ghosts.  Here the upper Latin index, $a$, denotes color and the lower, $b$, flavor. The gauge-covariant derivative $D_\mu$ and the $su(N)$ Lie bracket continue to act on the (upper) color index of $\phi^a_{\nu b}$ only.

The color and flavor indices of auxiliary ghosts both take values in the adjoint representations of an $su(N)$ algebra. Among other symmetries, the Lagrangian density ${\cal L}^{\rm gf}$ thus is invariant under separate global color and flavor transformations, on the upper and lower index respectively, of an  $SU(N)  \times SU(N)$ group.  With these specifications, the gauge-fixing term ${\cal L}^{\rm gf} = s \Psi$ of the Lagrangian density reads\footnote{In the following the dot-product is extended to include a summation over flavor when appropriate, $X\cdot Y\equiv \sum_{ab}X^a_b Y^a_b$. We also introduce the diagonal trace $\Tr X\equiv\sum_a X^a_a$,  and denote the adjoint component of an auxiliary ghost in the diagonal $su(N)$ subalgebra by $f^a[X]\equiv  \sum_{bc} g f^{abc} X^b_c$.}, 
\beqa
\label{unshiftedS}
\Psi & \equiv &  i \p_\mu \hat{\bar c}\cdot  A_\mu + \p_\mu \bar\omega_\nu\cdot D_\mu \phi_\nu,
\nonumber  \\
s \Psi & = & i \p_\mu \hat b\cdot A_\mu  - i \partial_\mu \hat{\bar c}\cdot  D_\mu c
+ \p_\mu \bar\phi_\nu \cdot D_\mu \phi_\nu 
- \p_\mu \bar\omega_\nu\cdot (D_\mu \omega_\nu + D_\mu c \times \phi_\nu) \ .
\label{sPsi}
\eeqa

Consider the change of variables introduced in~\cite{Schaden:1994},
\beqa
\label{MS}
\phi^a_{\nu b}(x) & = & \varphi^a_{\nu b}(x) - \gamma^{1/2} x_\nu \delta^a_b
\nonumber  \\
\bar\phi^a_{\nu b}(x) & = & \bar\varphi^a_{\nu b}(x) + \gamma^{1/2} x_\nu \delta^a_b  
\nonumber  \\
\hat b^a(x) & = & b^a(x) + i \gamma^{1/2} x_\nu f^a[ \bar\varphi_\nu(x)]
\nonumber  \\
\hat{\bar c}^a(x) & = & \bar c^a(x) + i \gamma^{1/2} x_\nu f^a[\bar\omega_\nu(x)],
\eeqa
 all other fields remaining the same.  Here $\gamma$ is a positive parameter whose value will be determined shortly.  This shift of the fields breaks the $SU(N)\times SU(N)$ color-flavor symmetry to a diagonal $SU(N)$ subgroup.  Remarkably this $x$-dependent change of variables does not introduce an explicit $x$-dependence into the Lagrangian density which, in terms of the shifted fields,  is given by
\beqa
\label{SGZ}
{\cal L}(\varphi, \bar\varphi, b, \bar c, \gamma) & = & {\cal L}(\phi, \bar\phi, \hat b, \hat{\bar c}) = {1 \over 4} F_{\mu \nu}^2 + s \Psi\ ,
\nonumber \\
\text{with }\ \Psi & \equiv &  i \p_\mu \bar c\cdot A_\mu + \p_\mu \bar\omega_\nu\cdot D_\mu \varphi_\nu - \gamma^{1/2}\Tr D_\mu \bar\omega_\mu\ ,
\nonumber  \\
{\cal L}^{\rm gf}=s \Psi & = & i \p_\mu b\cdot A_\mu  - i \partial_\mu \bar c\cdot D_\mu c
+ \p_\mu \bar\varphi_\nu\cdot D_\mu \varphi_\nu 
- \p_\mu \bar\omega_\nu\cdot (D_\mu \omega_\nu  + D_\mu c \times \varphi_\nu)
\nonumber \\
 &&+ \gamma^{1/2}\Tr [ D_\mu ( \varphi_\mu - \bar\varphi_\mu) - D_\mu c \times \bar\omega_\mu] - \gamma d (N^2 -1)\ . 
\eeqa
By \eqref{MS}, the BRST operator acts on the new fields according to
\begin{align}
\label{sonshiftedfields}
s A^a_\mu  &= (D_\mu c)^a &  s c^a &= - {1\over 2} (c \times c)^a
\nonumber   \\
s \bar c^a & = b^a &   s  b^a &= 0
\nonumber \\
s \varphi^a_{\mu b} &= \omega^a_{\mu b}  &   s \omega^a_{\mu b} &= 0 \ 
\nonumber \\
s \bar\omega^a_{\mu b} &=   \bar\varphi^a_{\mu b} + \gamma^{1/2} x_\mu \delta^a_b &   s \bar\varphi^a_{\mu b} &= 0.
\end{align}

\section{Poincar\'e Algebra}

The generator of a space-time translation of the unshifted fields is given by
\beq
{\cal P}_\nu = \int d^dx \ \mathfrak{p}_\nu
\eeq
\beqa
\mathfrak{p}_\nu = \p_\nu A_\mu \cdot {\delta  \over \delta A_\mu} + \p_\nu \hat b \cdot  {\delta \over \delta  \hat b}  + \p_\nu c \cdot  {\delta \over \delta c} 
  + \p_\nu {\hat{\bar c}} \cdot {\delta \over \delta {\hat{\bar c}}} +  \p_\nu \phi_\mu \cdot {\delta \over \delta \phi_\mu}  
 \nonumber \\
+ \p_\nu \bar\phi_\mu \cdot {\delta \over \delta \bar\phi_\mu}  + \p_\nu \omega_\mu \cdot  {\delta \over \delta \omega_\mu}  + \p_\nu \bar\omega_\mu \cdot  {\delta \over \delta \bar\omega_\mu},
\eeqa
as one sees by inspection.  In the unshifted action, \eqref{unshiftedS}, the unshifted auxiliary ghosts, such as $\phi_{\mu b}^a = \phi_B^a$, may be transformed under Lorentz transformation either as scalars or as vectors because both are symmetries of the action.  To be definite, we choose scalars and accordingly
\beq
{\cal M}_{\lambda \mu} = \int d^dx \ \left(x_\lambda \mathfrak{p}_\mu - x_\mu \mathfrak{p}_\lambda + A_\mu {\delta \over \delta A_\lambda} -  A_\lambda {\delta \over \delta A_\mu} \right).
\eeq
The last term effects the Lorentz transformation on the vector indices of $A_\nu$.  These operators satisfy the Poincar\'e commutation relations
\beq
\label{commutePoincare}
[ {\cal P}_\mu, {\cal P}_\nu ] = 0;  \ \ \ \ \ \  [ {\cal M}_{\lambda \mu}, {\cal P}_\nu ] = \delta_{\lambda \nu} {\cal P}_\mu - \delta_{\mu \nu} {\cal P}_\lambda
\eeq
\beq
 [ {\cal M}_{\lambda \mu}, {\cal M}_{\sigma \tau} ] = \delta_{\lambda \sigma} {\cal M}_{\mu \tau} - \delta_{\mu \sigma} {\cal M}_{\lambda \tau} - \delta_{\lambda \tau} {\cal M}_{\mu \sigma} + \delta_{\mu \tau} {\cal M}_{\lambda \sigma}.
\eeq
They are manifest symmetries of the action \eqref{unshiftedS} which is expressed in terms of the unshifted fields,
\beq
[ {\cal P}_\mu, S] = [ {\cal M}_{\lambda \mu}, S] = 0,
\eeq
Moreover they commute with the BRST charge,
\beq
\label{commuteQB}
[ Q_B, {\cal P}_\nu ] = [Q_B, {\cal  M}_{\lambda \mu}  ] = 0,
\eeq
where
\beq
\label{unshiftedforQB}
Q_B = \int d^dx \ \left[ D_\mu c\cdot {\delta \over \delta A_\mu} - \half (c \times c) \cdot {\delta \over \delta c} + \hat b\cdot {\delta \over \delta \hat{\bar c}} + \omega_\mu\cdot {\delta \over \delta \phi_\mu} + \bar\phi_\mu \cdot {\delta \over \delta \bar\omega_\mu} \right]\ .
\eeq
Altogether $\cal P_\nu$ and $\cal M_{\lambda \mu}$ have all the properties desired of physical Poincar\'e generators.\footnote{The action (\ref{SGZ}) is also invariant under Poincar\'e transformations of the shifted fields.  They define a second Poincar\'e symmetry algebra \cite{II}.}

\section{The variational vacuum}\label{newvacuum}

We look for a  vacuum in which the new fields have vanishing expectation value,
\beq
\label{shiftedvacuum}
\vev{\varphi^a_{\mu b}(x)} = \vev{\bar\varphi^a_{\mu b}(x)} = 0,
\eeq
and thus are well-behaved at $x = \infty$.  There should not be a new free parameter $\gamma$ in QCD.  To determine $\gamma$ we recall that the quantum effective action $\Gamma(\hat\Phi)$ is stationary at the vacuum configuration,
\beq
\delta\Gamma = \int d^dx \ { \delta \Gamma(\hat\Phi) \over \delta \hat\Phi_i(x) } \delta \hat\Phi_i (x) = 0,
\eeq  
for arbitrary infinitesimal variations $\delta \hat\Phi_i(x)$. Here $\hat\Phi_i(x)$ is the set of all the original elementary fields and their variation is unconstrained in that it need not vanish for $|x|\rightarrow\infty$.  In \eqref{MS}, the change of variables, which we write as $\hat\Phi = \hat\Phi( \Phi, \gamma)$, replaces the original unconstrained fields $\hat\Phi_i(x)$ by new fields $\Phi_j(x)$ that vanish at large $|x|$ and a variational parameter $\gamma$. Infinitesimal variations $\delta\hat\Phi$ of the old unconstrained fields amount to variations $\delta \Phi$ of the new constrained fields and variations $\delta \gamma$ of the parameter $\gamma$.  The new classical vacuum should be a minimum of the quantum effective action and is determined by the condition that the transformed quantum effective action $\Gamma( \Phi,  \gamma) = \Gamma(\hat\Phi)$ be stationary under these variations
\beq
\label{Gammastationary}
\delta \Gamma( \Phi,  \gamma) = \int d^dx { \delta \Gamma( \Phi,  \gamma) \over \delta \Phi_i } \delta \Phi_i + { \p\Gamma( \Phi,  \gamma) \over \p \gamma } \delta\gamma = 0.
\eeq
The quantum effective action $\Gamma( \Phi,  \gamma)$ can be calculated from ${\cal L}(\Phi, \gamma)$.  For a stationary point at $\Phi_i = 0$, \eqref{Gammastationary} reduces to the condition,
\beq
\label{horizoncondition}
 0 = { \p \Gamma \over \p \gamma } = -  {\p W \over \p \gamma } = \vev{ \p S \over \p \gamma } ,
\eeq
where $W$ is the free energy, and $S = S^{\rm YM} + \int d^dx\,  s \Psi$. By \eqref{SGZ}, the explicit form of \eqref{horizoncondition} becomes,
\beq\label{eqforgamma} 
\half\gamma^{1/2}\vev{\Tr (D_\mu (\varphi_\mu - \bar\varphi_\mu) - D_\mu c \times \bar\omega_\mu )} = \gamma d (N^2 -1)\ ,
\eeq
where the Euclidean space-time volume has been factored out. In Sect.~\ref{traceanomaly} we will see that the vacuum with $\gamma>0$ is energetically favored. 

We establish that the term in $c-\bar\omega$ of \eqref{eqforgamma} does not contribute\footnote{Perturbatively this is due to the absence of a $\omega-\bar c$ term in the GZ action.}. The $c-\bar\omega$ propagator in fact vanishes for any fixed gauge field $A_\mu^a(x)$,
\beq
\label{comegabarprop}
\vev{ c(x) \bar\omega(y) }_A = 0,
\eeq
where the restricted expectation value in the background $A$ is calculated by integrating over all fields except the gauge connection~$A$.   \eqref{comegabarprop} is a consequence of the phantom symmetry generated by the charge $Q_{R,\mu a}$ given in \eqref{QR}. Assuming this phantom symmetry is not spontaneously broken by the new vacuum, we have
\beq
0=\vev{ [Q_{R,\mu c}, c^b(x) \bar c^a(y) ] }_A = i \vev{c^b(x)\bar\omega^a_{\mu c}(0)}_A\ ,
\eeq
which gives \eqref{comegabarprop}. Dropping the (vanishing) $c-\bar \omega$-term in~\eqref{eqforgamma} and integrating out all the fields except the (transverse) gauge connection, one obtains
\beq
\label{horizonconditionb}
\int d^dy \ \vev{ D_\mu^{(x)ab} D_\mu^{(y)ac}(M^{-1})^{bc}(x, y; A)} = d (N^2 -1).
\eeq
Here $M \equiv - D_\mu \p_\mu$ is the Faddeev-Popov operator. This equation is equivalent to \eqref{horizonconditiona} because $i (M^{-1})^{ab}(x,y)=\vev{c^a(x) \bar c^b(y)}$ is the ghost propagator.   \eqref{eqforgamma} determines $\gamma$ or, more precisely, the ratio $\gamma / \Lambda_{QCD}^4$.  The equivalent \eqref{horizonconditionb} was originally derived~\cite{Zwanziger:1989mf,Zwanziger:1993} as the (horizon) condition that ensures positivity of the functional measure. 

\section{Analysis of BRST breaking}\label{BRSTbreaking}

\subsection{BRST lost}
\label{LostBRST}

The vacuum appears to break BRST symmetry spontaneously, for from \eqref{sonshiftedfields} we have
\beq\label{sbomega}
\vev{s \bar\omega^a_{\mu b}} = \vev{\bar\varphi^a_\mu+ \gamma^{1/2} x_\mu \delta^a_b} = \gamma^{1/2} x_\mu \delta^a_b\ .
\eeq
If we assume the existence of a well-defined BRST charge $Q_B$ that effects the $s$-operation, $\{ Q_B, \bar\omega^a_{\mu b}\} = s \bar\omega^a_{\mu b}$,  and a vacuum state $\vac$, the expectation value  $\vevv{\{ Q_B, \bar\omega^a_{\mu b}(x) \}} \neq 0$ formally implies that $Q_B \vac\neq 0$.  Here $Q_B$ is the BRST charge that in terms of the original fields is given in \eqref{unshiftedforQB},
and in terms of the new fields by,
\beq
\label{QB}
Q_B = \int d^dx \ \left[ D_\mu c\cdot {\delta \over \delta A_\mu} - \half (c \times c)\cdot {\delta \over \delta c} + b \cdot{\delta \over \delta \bar c} + \omega_\mu\cdot {\delta \over \delta \varphi_\mu} + \bar\varphi_\mu\cdot {\delta \over \delta \bar\omega_\mu} + \gamma^{1/2} x_\mu \Tr {\delta \over \delta \bar\omega_\mu} \right]\ .
\eeq
This follows from \eqref{MS}, which implies that the partial derivatives transform according to,
\begin{align}
\label{shiftpartials}
{\delta \over \delta \phi^a_{\mu b} } & \to  {\delta \over \delta \varphi^a_{\mu b} } &  
{\delta \over \delta \bar\phi^a_{\mu b} } &\to {\delta \over \delta \bar\varphi^a_{\mu b} } - i \gamma^{1/2} x_\mu f^{abc} {\delta \over \delta b^c }
\nonumber \\
{\delta \over \delta \omega^a_{\mu b} } & \to {\delta \over \delta \omega^a_{\mu b} } & {\delta \over \delta \bar\omega^a_{\mu b} } &\to {\delta \over \delta \bar\omega^a_{\mu b} } - i \gamma^{1/2} x_\mu f^{abc} {\delta \over \delta \bar c^c }
\nonumber  \\ 
{\delta \over \delta\hat b^a } & 
\to {\delta \over \delta  b^a }& {\delta \over \delta \hat{\bar c}^a } &\to {\delta \over \delta \bar c^a }.
\end{align}

\subsection{The surface equation}
\label{surface equation}

The spontaneous breaking of BRST symmetry in \eqref{sbomega} is puzzling at first for the action $S$ is BRST-invariant, $sS = 0$. The breaking therefore takes the form
\beq
\int d\Phi \ s[{\bar \omega}^a_{\mu b}\, \exp(- S)  ] \neq 0,
\eeq
where $s$ is the fermionic derivative given by \eqref{QB}, whereas the integral of any well-defined fermionic derivative should vanish, $\int d\Phi \ sG(\Phi) = 0$, as one sees in a mode expansion.    

To resolve this paradox, we quantize in a finite volume $L^d$.  We impose periodic boundary conditions in every Euclidean direction $\mu$, $\Phi(x_\mu + L ) = \Phi(x_\mu)$, on all fields, $\Phi = (A, c, \bar c,  b, \varphi, \bar\varphi, \omega, \bar\omega)$, that appear in the shifted action of \eqref{SGZ}.  The operator $s$, introduced in \eqref{sonshiftedfields}, is no longer well-defined, because the function $x_\mu$ is not periodic.  We instead introduce an operator $s_L$ that is compatible with the periodic boundary conditions and defines the $s$-operator when the boundary recedes to infinity.

To this end, we introduce the periodic saw-tooth function,
\begin{align}
\label{sawtooth}
h(x_\mu) & = x_\mu\ ,  {\rm for}\  - L/2 < x_\mu < L/2
\nonumber \\
h(\pm L/2)&=0\ ;\  h(x_\mu + L) =  h(x_\mu)\ ,
\end{align}
which agrees with the linear function $x_\mu$ for $ -L/2 < x_\mu < L/2$, and has the derivative,
\beq
\label{hprime}
\p_\nu h(x_\mu) = \delta_{\mu \nu} \left(1 - \frac{L}{2} \delta(x_\mu - L/2 )-\frac{L}{2} \delta(x_\mu + L/2 )\right)  \ {\rm for} \  - L/2 \leq x_\mu \leq L/2,
\eeq
The sawtooth has a vertical stroke of length $L$ that we have placed at the boundary of the interval.  At the end of the day we shall take the infinite-volume limit $L \to \infty$.

Let us now define an  operator $s_L$ that is consistent with the periodic boundary conditions whose action on the fields is,
\beqa
\label{ssubL}
s_L \bar\omega^a_{\mu b}(x) & = & \bar\varphi^a_{\mu b}(x) + \gamma^{1/2} h(x_\mu) \delta^a_b
\nonumber  \\
s_L \Phi(x) & = & s\Phi(x) \ \ {\rm for} \ \ \Phi \neq \bar\omega\ .
\eeqa
$s_L$ is nil-potent, $s_L^2 = 0$.  Although not a symmetry of the action, $s_L S \neq 0$, this fermionic derivative has the advantage of being well-defined.  At interior points $y$ of the quantization volume, the local Lagrangian density satisfies $s {\cal L}(y) = 0$,  so only the vertical stroke of the saw-tooth contributes to $s_L S = (s_L - s)S$, which by \eqref{SGZ} gives,
\beq
\label{surfacebreaking}
s_LS = \gamma^{1/2} \frac{L}{2}\sum_{\mu =1}^d\sum_{\sigma=\pm} \int_{x_\mu=\sigma L/2}\hspace{-2em} dS_\mu  \Tr [ D_\mu \omega_\mu + D_\mu c \times \varphi_\mu ](x),
\eeq
where the integral extends over the surfaces at $ x_\mu=\pm L/2$.  The breaking of $s_LS$ is here expressed as an integral over the boundary of the elementary hypercube.

Due to this explicit breaking, it is not true that $\vev{ s_L F}$ vanishes for every operator $F$.  Instead we have
\beq
\int d \Phi \  s_L  [F\, \exp(-S) ] = 0,
\eeq
because the integral of a {\em well-defined} fermionic derivative vanishes.  This gives
\beq
\vev{s_L F} = \vev{F\, s_L S}.
\eeq
Suppose $F$ is concentrated at points $y$ that are in the interior of the quantization volume $|y| < L/2$, well away from the vertical stroke of the saw-tooth function. In this case\footnote{For simplicity, we take $F = F(y)$ to be concentrated at a single point $y$.},
\beq
s_L F(y) = s F(y) \hspace{2cm} |y_\mu| < L/2.
\eeq
  This gives
\beq
\vev{s F(y)} = \vev{F(y)\  s_L S},
\eeq
and with \eqref{surfacebreaking} one obtains, for $|y_\mu| < L/2 $,
\beq
\label{sLbreak}
\vev{s F(y)}= \gamma^{1/2} \frac{L}{2}\sum_{\mu =1}^d\sum_{\sigma=\pm} \int_{x_\mu=\sigma L/2}\hspace{-2em} dS_\mu \   \vev{F(y)\ \Tr [ D_\mu \omega_\mu + D_\mu c \times \varphi_\mu ](x)}.
\eeq
The breaking of BRST symmetry at a point $y$ in the interior of the quantization volume is here expressed as an integral at the surfaces with $x_\mu = \pm L/2$.   In Appendix~\ref{checksurfaceeq} we verify \eqref{sLbreak} by explicit calculation for the special case $F = \bar\omega^a_{\mu b}$.   An alternative expression for $\vev{s F }$ is provided in Appendix \ref{alternativemeansPi}.

\subsection{A sufficient condition for an operator to preserve BRST symmetry} \label{sufficientcondition}

We say an $s$-exact operator $sF$ breaks (or preserves) BRST symmetry if its vacuum expectation value is non-zero, $\vev{sF} \neq 0$, (or zero).  For functionals $F$ of physical interest we will need to determine whether $\vev{ s F} = 0$.  Although this is not true for certain operators, as for instance $\bar\omega^a_{\mu b}$, it is true that $\vev{ s F} = 0$ for a large class of local operators $F$.  From the surface equation we deduce a simple sufficient condition which assures that $\vev{ s F} = 0$.  

Consider the correlator $C_\mu(x)$ defined by$^7$,
\beq\label{corrF}
C_\mu(x - y) \equiv \langle \chi_\mu(x) F(y) \rangle,
\eeq
where
\beqa
\chi_\mu(x) & \equiv & s\Tr D_\mu \varphi_\mu(x) \hspace{2cm} {\rm (no \ sum \ over \ } \mu)
\nonumber \\
 & = &  \Tr ( D_\mu \omega_\mu + D_\mu c \times \varphi_\mu )(x).
\eeqa

{\em Sufficient condition theorem:}  If $C_\mu(x-y)$ satisfies,
\beq
\label{sufficient}
\lim_{|x| \to \infty} |x|^d \ C_\mu(x - y) =0, \ \text{for all directions} \ \mu,
\eeq
where $d$ is the dimension of Euclidean space-time, then $\vev{s F(y)} = 0$ in the infinite volume limit.    This condition requires the fall-off of the correlator $\langle \chi_\mu(x) F(y) \rangle$ to be sufficiently rapid at large separation.

The proof is immediate. Since the surface of the integration volume at $x_\mu=O(L/2)$ is of order $O(L^{d-1})$, we deduce from  \eqref{sLbreak} and \eqref{sufficient} that, 
 \beq
\label{ssubLbreak}
\langle s F(y) \rangle =\ \gamma^{1/2} \sum_\mu O(L^d C_\mu (L/2))\xrightarrow{L\rightarrow\infty} 0 .
\eeq

\section{BRST construction of the physical states}\label{BRST-GZconstruction}

\subsection{Definition of observables} \label{physoperators}

We have seen in Sect.~\ref{surface equation} that the BRST symmetry is spontaneously broken in the infinite-volume limit, because $\vev{s F} \neq 0$ for some operators, such as $s\bar\omega^a_{\mu b}$.  It would not be satisfactory if this occurred for an operator $s F$  in the class of observables.  For example, as discussed in the Introduction, one would like to replace the energy-momentum tensor $T_{\mu \nu} = T_{\mu \nu}^{\rm YM} + s \Xi_{\mu \nu}$ by the Yang-Mills energy-momentum tensor $T_{\mu \nu}^{\rm YM}$ in physical Ward identities.


There are a number of unphysical ghosts, and a rich class of symmetry transformations, with generators $Q_Y$ that act on ghost degrees of freedom only.  The index $Y$ here specifies the ghost symmetry.  For example, there is an obvious $SU(N)$ symmetry that acts on the lower (flavor) index of the unshifted auxiliary ghost fields according to $[ Q^c, \bar\phi^a_{\mu b}] = f^{bcd} \bar\phi^a_{\mu d}$, etc., where the conserved charge $Q^c$ generates the symmetry.\footnote{The symmetries of the unshifted action are easily recognized, and all symmetries of the unshifted action are symmetries of the shifted action when expressed in terms of the shifted fields.  The charge $Q^c$, written in terms of the shifted fields, is the linear combination  $Q^a=\half \sum_{\mu bc}f^{abc}Q_{F,\mu\mu bc}$ of charges $Q_{F,\mu\mu bc}$ defined  in \eqref{QF} of Appendix~\ref{phantomsymm}.}  How do the ghost symmetries help characterize observables?

A physical observable~$G$ of a gauge theory depends on gauge-invariant degrees of freedom only, as for example $G = G(F_{\mu \nu}^2, \bar\psi \psi)$, where $\psi$ is the quark field. It  therefore commutes with all generators $Q_Y$ of ghost symmetries, 
\beq
[Q_Y, G(F_{\mu \nu}^2, \bar\psi \psi)] = 0.
\eeq
Requiring observables of GZ-theory to commute with \emph{all} ghost symmetry generators $Q_Y$ serves to ensure that the class of observables is not larger than it should be in a gauge theory without infringing on any gauge-invariant functional.  Of course, we also require observables~$G$ to be BRST-invariant $[Q_B, G] = 0$.

Accordingly the class ${\cal W}_{\rm phys}$ of Euclidean observables of GZ-theory is defined by
\beq
\label{physicalops}
{\cal W}_{\rm phys} \equiv \{G:  [Q_B, G] = [Q_Y, G] =0, \text{ for all ghost symmetries } Q_Y\},
\eeq
where $G = G(\Phi)$ is a local polynomial in the elementary Euclidean fields $\Phi_i = (A, c, \bar c, b, \varphi, \bar\varphi, \omega, \bar\omega)$.  An immediate consequence of this definition is that, by the Jacobi identity, all graded commutators,  $Q_Z = [Q_B, Q_Y]_\pm$, of $Q_B$ with a ghost symmetry $Q_Y$ also generate symmetries of the observables, $[Q_Z, G]_\pm = 0$, and we may equivalently define the class of observables by
\beq
\label{physicalopsa}
{\cal W}_{\rm phys} \equiv \{G:  [Q_X, G] =0, \text{ for all } Q_X\in\mathfrak{F}\},
\eeq
where $\mathfrak{F}$ is the set of generators in the closed algebra containing $Q_B$ and the ghost charges $Q_Y$.  The set $\mathfrak{F}$ is given in \eqref{Fset}.  Since  these symmetries leave all observables invariant, they cannot be observed and we call them phantom symmetries. 
 
The generators of phantom symmetries in terms of shifted fields are collected in Appendix~\ref{phantomsymm}. The set $\mathfrak{F}$ includes the BRST charge $Q_B$ and the ghost number $Q_{\cal N}$, but the closed algebra of unphysical charges in the GZ theory  is much larger.  Note that the generator of (unbroken) rigid color transformations, $Q^a_C=[Q_B, Q^a_G]$, is part of a BRST doublet, where $Q_C^a$ and $Q_G^a$ are given by \eqref{QC} and \eqref{QG}. This is a feature of Landau gauge~\cite{Blasi:1991} and not peculiar to the GZ theory. The global color charge $Q_C$ thus is the BRST-variation of a ghost symmetry and is included in the closed algebra of phantom symmetries $\mathfrak{F}$,
\beq
\mathfrak{F} = \{ Q_X \} = \{ Q_B, Q_C, Q_Y \},
\eeq
where $Q_X$ is the set of all phantom symmetries, $Q_Y$ is the set of symmetries that act on the ghost variables only and $Q_C$ generates global color transformations.  Observables in this sense are color singlets. 

\subsection{BRST regained}\label{hypothesis}  

The annoying operator $s\bar\omega$, with $\vev{ s \bar\omega} \neq 0$ is \emph{excluded} from the class ${\cal W}_{\rm phys}$ of observables, because, among other phantom symmetries,  $[Q_{\bar\varphi,\mu}, s\bar\omega^a_{\nu b}] = \delta^a_b \delta_{\mu\nu} \neq 0$, where $Q_{\bar\varphi,\mu}$ is the phantom symmetry generator of \eqref{Qphibar}.  In fact the condition that $[Q_X, sY] = 0$ for all phantom symmetries $Q_X$ is quite restrictive for $s$-exact observables $G = sY$, as a look at Appendix \ref{phantomsymm} reveals. However,  ${\cal W}_{\rm phys}$ does include some $s$-exact observables such as $s\Psi$ given in \eqref{SGZ} because, by definition, phantom symmetries are symmetries of the Lagrangian density. As shown below, the $s$-exact part of the energy momentum tensor, $T_{\mu \nu} = T_{\mu nu}^{\rm YM} + s\Xi_{\mu \nu}$ is another. 

  If the expectation value of {\em every} $s$-exact functional in ${\cal W}_{\rm phys}$ vanishes ($\vev{sY} = 0$ for $sY \in {\cal W}_{\rm phys}$), the physical state space reconstructed from the correlators $\vev{G(\Phi)}$ with $G(\Phi)\in{\cal W}_{\rm phys}$ would enjoy an unbroken BRST symmetry.  Although we cannot prove that this is the case, neither have we found evidence to the contrary. The example of Appendix~\ref{applycriterion} shows that the class of BRST-exact functionals with vanishing expectation value is in fact not limited to $\Wphys$. However, in many cases it is difficult to verify whether $\vev{s \Sigma}$ vanishes or not, because this generally  depends on the non-perturbative horizon condition.  Where we could do the calculation, we found that the expectation values of the $s$-exact parts of the energy-momentum tensor and of the Lagrangian as well as (to leading order) the $s$-exact term in the Kugo-Ojima equation indeed vanish curtesy of the horizon condition.  In view of the above considerations we shall take as a\vspace{3ex}

\noindent{\em Hypothesis}:  BRST symmetry is not broken by $s$-exact observables,
\beq
\vev{sY}= 0 \ {\rm for \ all} \ sY\in {\cal W}_{\rm phys},
\eeq
where ${\cal W}_{\rm phys}$ is the set of observables defined in \eqref{physicalops}.

\subsection{BRST cohomology and physical states}
\label {cohomology}

Provided the hypothesis holds, BRST symmetry is unbroken by the observables and all conditions for reconstructing  the physical space of a gauge theory are satisfied in the GZ model.\footnote{BRST symmetry holds order by order in FP theory, but it is a hypothesis that it remains unbroken non-perturbatively.}    We suppose that the vacuum expectation values $\vev{F(\Phi) }$ of all local polynomials $F(\Phi)$ are given, and the physical Euclidean state space will be reconstructed from these correlators.

  Physical observables form a vector space under addition: if $F_1\text{ and } F_2 \in \Wphys$, then $F = c_1F_1 + c_2 F_2 \in \Wphys$.  This vector space is provided with an inner product,  
\beq
\braket{F}{G} \equiv \vev{ F^\dag G} \  {\rm for} \  F, G \in {\cal W}_{\rm phys},
\eeq
where the hermitian conjugate of the fields is given by,\footnote{The minus signs could be avoided by the replacements $ib = b', i \bar c = \bar c'$. }  
\beq
A^\dag = A^\dag; \ \  b^\dag = - b; \ \  c^\dag = c; \ \  {\bar c}^\dag = - {\bar c}; \ \  \varphi^\dag = \varphi; \ \  \bar\varphi^\dag = \bar\varphi; \ \  \omega^\dag = \omega; \ \  \bar\omega^\dag = \bar\omega.
\eeq
We define a (Euclidean) pre-physical state to be an observable $F \in \Wphys$ which, to emphasize its vector property, we also designate by $\ket{F}$.  Pre-physical states that are $s$-exact $\ket{s \Xi}$ form a linear subspace ${\cal W}_0 \subset \Wphys$,
\beq
{\cal W}_0 \equiv \{ sY: sY \in \Wphys \}.
\eeq

\vspace{2ex}
\noindent{\em Lemma:}  Every pre-physical state in ${\cal W}_0$  is orthogonal to all pre-physical states
\beq
\braket{ F}{sY} = 0 \text{ for all } F \in \Wphys \text{ and } sY \in {\cal W}_0.
\eeq
The above hypothesis indeed implies that,
\beq
\label{orthogonal}
\braket{F}{s Y} = \vev{ F^\dag s Y} 
= \vev{s( F^\dag Y) } = 0,
\eeq
where we have used the fact that $\Wphys$ is closed under hermitian conjugation and multiplication, so $F^\dag s Y \in {\cal W}_{\rm phys}$, and that $sF^\dag = 0$ for $F^\dag \in {\cal W}_{\rm phys}$.  Thus ${\cal W}_0$ is a null subspace of $\Wphys$.  

In the following section we show that the Euclidean inner product is positive and we define the (Euclidean) physical Hilbert space to be the completion in the norm of the quotient space,
\beq
{\cal H}_{\rm phys} = \overline{ \Wphys / {\cal W}_0 }.
\eeq
Physical states are thereby associated to the BRST cohomology and are equivalence classes $\ket{ \{ G \} }$ of pre-physical states of the form $\ket{ G + s X}$, where $G, s X \in \Wphys$, and $G$ is not $s$-exact $G \neq sY$.

The BRST operator acts trivially on the unshifted auxiliary ghosts in \eqref{sonaux} and the cohomology of the GZ theory therefore is the same as that of Faddeev-Popov theory.  The proofs in~\cite{Dixon:1977,Barnich:2000zw} that the cohomology is free of the unshifted BRST doublets carry over to the shifted fields because the MS-shift is an invertible linear transformation of the doublets.  Consequently, every equivalence class $\{G\}$ has a representative $G(A)$ that is a gauge-invariant functional of the gauge connection $A$ only:
\beq
\label{GisGofA}
G(A) \in \{ G \}, \ \ {\rm and} \ \  s G(A) = 0.
\eeq

\subsection{Positivity of the Euclidean inner product}\label{positivity}

The Euclidean inner product on the space of physical states is defined by
\beq
\langle F | G \rangle = { \int d\Phi \ \exp(- S ) F^\dag G \over \int d\Phi \ \exp(- S )}.
\eeq
It is essential that this inner product be non-negative for $F = G$. This is precisely what the GZ action was designed~\cite{Zwanziger:1989mf} to do, as we now recall.

The original motivation for the present approach was to impose a cut-off at the Gribov horizon because every gauge orbit passes inside the Gribov region \cite{Dell'Antonio:1991}.  That led directly to the non-local action (\ref{nonlocal}) which can be reexpressed as the local GZ action.  In the present article we have followed the alternative derivation of the GZ action \cite{Schaden:1994}, which is based on the MS shift (\ref{MS}) and makes no reference to eliminating Gribov copies.  In this approach, the shifted fields themselves may be said to cut the functional integral off at the Gribov horizon, in the sense that the cut-off factor $\exp[- \gamma H(A)]$ that appears in (\ref{nonlocal}), below, has an essential singularity as the Gribov horizon is approached from the inside, and vanishes there together with all its derivatives \eqref{essntialsingularity}.  Any Green's function which is evaluated analytically, for example in a diagrammatic expansion, will only receive contributions from the interior of the Gribov region.  The theory is no longer defined outside this region.  

 By \eqref{GisGofA} we may choose as representative of any physical state, a (gauge-invariant) functional that depends on the connection $A$ only, $F = F(A)$ and $G = G(A)$.  One then can integrate out the Lagrange multiplier field $b$. This imposes the gauge condition and restricts the functional integral to transverse connections, $\p \cdot A = 0$, for which the Faddeev-Popov operator $M(A)$ is hermitian.  Next one integrates out the Faddeev-Popov ghosts, which gives the Faddeev-Popov determinant $\det [ M(A)] = {\prod_n}' \lambda_n(A)$, where $\lambda_n(A)$ are the eigenvalues of $M(A)$, and the prime indicates that the trivial null eigenvalues due to rigid gauge transformations are to be excluded.  This determinant is positive inside the Gribov region by definition, for that is the region where all (non-trivial) eigenvalues are positive.  We finally integrate out the auxiliary ghosts by Gaussian integration.  This results in a cut-off factor, $ \exp[ - \gamma H(A) ] $, where
\beq
H(A) \equiv g^2  \int d^dx d^dy \ f^{abc} A_\mu^b(x) (M^{-1})^{cd}(x, y) f^{aed} A_\mu^e(y) 
\eeq
is the ``horizon function", and $(M^{-1})^{cd}(x, y)$ is the kernel of the inverse Faddeev-Popov operator.  Only the $A$-integration remains,
\beq
\label{nonlocal}
\langle F | G \rangle = N \int_{\partial \cdot A=0} \hspace{-1em}dA \ \det [ M(A)]\exp[- \gamma H(A)] \ F^*(A) G(A).
\eeq
It was shown in~\cite{Zwanziger:1989mf}, by an argument similar to the proof of the equivalence of the micro-canonical and canonical ensembles in statistical mechanics, that a sharp cut-off at the boundary of the Gribov region $\Omega$ is equivalent to the cut-off factor $\exp[ - \gamma H(A)]$, provided that $\gamma$ has the value determined by the horizon condition of \eqref{horizonconditionb}.  Thus the Euclidean inner product is equivalent to
\beq
\langle F | G \rangle = N \int_\Omega dA \ \det[M(A)] \ F^*(A) G(A)\ .
\eeq
This is a positive inner product, $\langle F | F \rangle \geq 0$, on the physical space of (gauge-invariant) functionals.   The calculations reported in the present article indicate that BRST may be preserved in the physical sector, and that the GZ action may provide a consistent quantization of a gauge theory when the horizon condition holds.  
    
In this context the behavior of the cut-off function $\exp[ - \gamma H(A)]$ as the Gribov horizon is approached is of interest.  For a given  configuration of the transverse gauge field $A$, the spectral representation of the inverse Faddeev-Popov operator in terms of eigenfunctions and eigenvalues,
\beq
(M^{-1}(A))^{ab}(x, y) = {\sum_n}' \ { u_n^a(x;A) u_n^b(y;A) \over \lambda_n(A) },
\eeq
gives
\beq
\exp[ - \gamma H(A) ] = \exp \left[ - \gamma {\sum_n}' \  { c_n^2(A) \over \lambda_n(A) } \right],  
\eeq
where $c_n^2(A) = \sum_{\mu a} c_{n \mu a}^2(A)$, with amplitudes
\beq
c_{n \mu a} = \int d^dx \ (A_\mu \times u_n)^a.
\eeq
The (non-trivial) eigenvalues $\lambda_n(A)$ are positive when $A$ is in the interior of the Gribov region $\Omega$, and the lowest non-trivial eigenvalue  approaches zero, $\lambda_0(A) \to 0^+$, as $A$ approaches the Gribov horizon $A \to \p \Omega$. For configurations near the Gribov horizon,  $H(A) \sim c_0^2(A) / \lambda_0(A)$.  The cut-off function thus has an essential singularity as the Gribov horizon is approached, where it vanishes, together with all its derivatives,
\beq
\label{essntialsingularity}
\lim_{A \to \p \Omega} \exp[ - \gamma H(A) ] \sim \lim_{\lambda_0 \to 0} \exp[ - \gamma c_0^2 / \lambda_0 ] = 0.
\eeq
This analysis suggests that the cut-off function $\exp[ - \gamma H(A) ]$ cuts off the integral at the Gribov horizon for any positive value of $\gamma$. However, only the value of $\gamma$ selected by the horizon condition of \eqref{horizonconditionb} may preserve the BRST symmetry of the physical space, as we demonstrate below.

\section{A non-trivial test: the energy-momentum tensor}\label{testcase}

\subsection{Derivation of the energy-momentum tensor}
\label{EMtensor}

In Faddeev-Popov theory the energy-momentum tensor is,
\beq
T_{\mu \nu}^{\rm FP} = T_{\mu \nu}^{\rm YM} + s( i \p_\mu \bar c\cdot A_\nu+  i \p_\nu \bar c\cdot A_\mu - \delta_{\mu \nu}  i \p_\lambda \bar c\cdot A_\lambda) , 
\eeq  
with $T_{\mu \nu}^{\rm YM}$ given by \eqref{TYM}.  It is symmetric, $T_{\mu \nu}^{\rm FP} = T_{\nu \mu}^{\rm FP}$,  and conserved, $\p_\mu T_{\mu \nu}^{\rm FP} = 0$, modulo the equations of motion.

To obtain the energy-momentum tensor of the GZ theory, we consider the action of \eqref{totallagrangian} written in terms of unshifted fields, $\hat\Phi$, without specifying the index set $B$.   Treating the auxiliary ghosts as scalar fields, we follow standard procedure and write the action $S = S(\hat\Phi, g)$ on a Riemannian background with metric $g_{\mu \nu}$.  The internal phantom symmetry generators $Q_X$ of Appendix~\ref{phantomsymm}, including the BRST charge $Q_B$, are also symmetries of the action on an arbitrary Riemannian background,\footnote{The generators of  Appendix~\ref{phantomsymm} are given for arbitrary values of $\gamma$. One obtains the generators for the unshifted variables at $\gamma=0$. At $\gamma=0$ none of the phantom symmetry generators is explicitly coordinate dependent.} $[Q_X, S(\hat\Phi, g)] = 0$.  It follows that they are symmetries of the functional derivative $T_{\mu \nu} =  \delta S(\hat\Phi, g) / \delta g_{\mu \nu} $ which is the energy-momentum tensor $[Q_X, T_{\mu \nu}] = 0$.  The symmetric, conserved energy-momentum tensor corresponding to the action of \eqref{totallagrangian} one obtains in this manner has the form,
\beq
\label{totalT}
T_{\mu \nu} = T_{\mu \nu}^{\rm YM} + s \Xi_{\mu \nu}.
\eeq
where
\beq
\Xi_{\mu \nu} =  i \p_\mu \hat {\bar c}\cdot A_\nu + \p_\mu \bar\omega_B\cdot D_\nu \phi_B + [\mu \leftrightarrow \nu] - \delta_{\mu \nu} \Psi ,
\eeq
with $\Psi$ is given in \eqref{PsiB}.  The Yang-Mills energy-momentum tensor is separately invariant under all phantom symmetries, which implies that $s \Xi_{\mu \nu}\in{\cal W}_{\rm phys}$ as well,
\beq
[Q_X, s \Xi_{\mu \nu}] = 0\ ,
\eeq 
for all $Q_X$, including $Q_B$. We shall express $\Xi_{\mu\nu}$ in terms of the shifted fields that are well defined at large $|x|$.  Let us first reintroduce the index set $B = (b, \kappa)$ to be a flavor index $b$ and (in flat space) a Lorentz index $\kappa$,
\beq\label{Xi}
\Xi_{\mu \nu} = [ i \p_\mu \hat{\bar c}\cdot A_\nu + \p_\mu \bar\omega_\kappa\cdot D_\nu \phi_\kappa ] + [\mu \leftrightarrow \nu] - \delta_{\mu \nu}(  i \p_\lambda \bar c\cdot A_\lambda + \p_\lambda \bar\omega_\kappa\cdot D_\lambda \phi_\kappa).
\eeq
Since nothing was done but give a name to the index set, the tensor $T_{\mu \nu}$ remains symmetric and conserved.

Next one performs the MS shift of \eqref{MS}.  It again works its magic and gives coordinate-independent tensors, 
\beqa\label{Xishifted}
\Xi_{\mu \nu} = [ i \p_\mu \bar c\cdot A_\nu + \p_\mu \bar\omega_\kappa\cdot D_\nu \varphi_\kappa - \gamma^{1/2} \Tr D_\mu \bar\omega_\nu] + [\mu \leftrightarrow \nu]
\nonumber \\
 - \delta_{\mu \nu}(  i \p_\lambda \bar c\cdot A_\lambda + \p_\lambda \bar\omega_\kappa\cdot D_\lambda \varphi_\kappa - \gamma^{1/2} \Tr D_\lambda \bar\omega_\lambda)
\eeqa
\beqa\label{sXishifted}
s\Xi_{\mu \nu} & = & [ i \p_\mu b\cdot A_\nu - i \p_\mu \bar c\cdot D_\nu c + \p_\mu \bar\varphi_\kappa\cdot D_\nu \varphi_\kappa  -  \p_\mu \bar \omega_\kappa\cdot (D_\nu \omega_\kappa + D_\nu c \times \varphi_\kappa)
\nonumber \\  \label{sXi}
&&+ \gamma^{1/2} \Tr (D_\mu \varphi_\nu
 - D_\mu \bar\varphi_\nu - D_\mu c \times \bar\omega_\nu) - \gamma (N^2-1) \delta_{\mu \nu} ] + [\mu \leftrightarrow \nu ] - \delta_{\mu \nu} {\cal L}^{\rm gf} 
\eeqa
where $\bar c, b, \varphi$ and $\bar\varphi$ are the shifted fields.  Here $s$ is the BRST-operator that acts on the shifted fields as in  \eqref{sonshiftedfields}. It is a symmetry of the Lagrangian density~${\cal L}$ of \eqref{SGZ}.  Because the MS-shift is but a change of variables, $T_{\mu \nu}$ remains conserved, modulo the equations of motion.




\subsection{Test of the hypothesis}
\label{testhypothesis}

We shall show that,
\beq
\label{vevTequalsvevTmunu}
\vev{ T_{\mu \nu} } = \vev{ T_{\mu \nu}^{\rm YM} }\ .
\eeq
This may be somewhat surprising, because the new vacuum breaks the symmetry of the bose and fermi ghosts which are transformed into each other by the BRST operator.  We here give a proof that relies on the sufficient condition of Sect.~\ref{sufficientcondition}.  In Appendix~\ref{testBRST} we provide a more direct, but perhaps less intuitive, alternative proof that uses the  equations of motion. Both methods require that the horizon condition of \eqref{horizonconditiona} be satisfied.   

We evaluate the vacuum-expectation value $\vev{s \Xi_{\mu \nu}}$, of the BRST-exact part of $T_{\mu \nu} = T_{\mu \nu}^{\rm YM} + s \Xi_{\mu \nu}$.  Exploiting Euclidean rotational symmetry one has,
$\vev{s \Xi_{\mu \nu} }= \delta_{\mu \nu} \vev{ s \Xi_{\lambda \lambda} } /d$, with $\vev{s \Xi_{\mu \mu} } = (2 - d) \vev{ s \Psi }$, where $\Psi$ is given in \eqref{SGZ}.  We rearrange the derivative and obtain
\beq
\vev{ s \Psi } = \vev{ s \Psi' } + \vev{ s \Psi'' }
\eeq  
\beq
\vev{ s \Psi' } = \vev{s \p_\mu [ i \bar c\cdot A_\mu + \p_\mu \bar\omega_\nu\cdot \varphi_\nu ]}
\eeq
\beq
\vev{ s \Psi'' } = \vev{ s [ -  i  \bar c\cdot \p_\mu A_\mu - D_\mu \p_\mu \bar\omega_\nu\cdot  \varphi_\nu - \gamma^{1/2}\Tr D_\mu \bar\omega_\mu ] }.
\eeq
Due to translation invariance we have $\vev { \p_\mu F } = 0$ for any local field $F$ that satisfies $[P_\mu, F] = \p_\mu  F$.  We cannot quite use this argument to argue that $\vev{ s \Psi' }$ vanishes because $s \bar\omega^a_{\mu b}=\bar\varphi^a_{\mu b}+x_\mu \gamma^{1/2} \delta_b^a$ depends explicitly on $x$.  This is the only term that could contribute to $\vev{ s \Psi' }$ in a translationally-invariant vacuum, and thus 
\beq
\vev{ s \Psi' } = \gamma^{1/2} \vev{ \p_\mu \Tr \varphi_\mu } = 0.
\eeq

To apply the  criterion of \eqref{sufficient} to $\vev{ s \Psi'' }$ we estimate the correlator,
\beq
\label{estimatecor}	
C_\lambda(x) = \vev{ \Tr ( D_\lambda \omega_\lambda + D_\lambda c \times \varphi_\lambda )(x) \ (  i  \bar c\cdot \p_\mu A_\mu + D_\mu \p_\mu \bar\omega_\nu\cdot  \varphi_\nu + \gamma^{1/2}\Tr D_\mu \bar\omega_\mu )(0)} \ .
\eeq
(no sum on $\lambda$) at large $|x|$.  We use the equation of motion of $b$ to set $\p_\mu A_\mu = 0$ and the previous result of \eqref{comegabarprop} that $\vev{ c(x) \bar\omega(y) }_A = 0$.  It follows that the term in $c$ vanishes in \eqref{estimatecor}, and we need only consider the asymptotic behavior of,
 \beq
\label{estimatecora}	
C_\lambda(x) = \vev{ \Tr  D_\lambda \omega_\lambda(x) \ ( D_\mu \p_\mu \bar\omega_\nu\cdot  \varphi_\nu + \gamma^{1/2}\Tr D_\mu \bar\omega_\mu) ](0)} \ .
\eeq
Integrating out the ghosts at fixed $A$ one has,
\beq
\omega^a_{\lambda b}(x) \bar\omega^c_{\mu d}(y) \to - (M^{-1})^{ac}(x, y; A) \delta_{bd} \delta_{\lambda \mu },
\eeq
which leads to,
\beqa
C_\lambda(x) = \vev{ D_\lambda^{(x)ab} \delta^d(x) \varphi^a_{\lambda b}(0) - \gamma^{1/2} D_\lambda^{(x)ab} D_\lambda^{(y)ac}(M^{-1})^{bc}(x, 0)}\ .
\eeqa
The first term vanishes, because for $|x| = O(L)$,  $\delta^d(x) = 0$.  One thus finds,
\beq
C_\lambda(x) = - \gamma^{1/2} \left.\vev{D_\lambda^{(x)ab} D_\lambda^{(y)ac}(M^{-1})^{bc}(x, y)}\right|_{y=0}\ .
\eeq
The asymptotic behavior for large $|x|$ of this correlator was obtained in Appendix~\ref{applycriterion}, and the sufficient condition of Sect.~\ref{sufficientcondition} implies that  $\vev{s \Xi_{\mu \nu} } =\vev{s \Psi} =0$. We thus could verify the hypothesis that BRST-exact observables have vanishing expectation value for this case and conclude that \eqref{vevTequalsvevTmunu} indeed holds in the GZ-theory.

  \section{Tree-level evaluation of the trace anomaly}\label{traceanomaly}
 
 Having found that $\langle T_{\mu \nu} \rangle = \langle T_{\mu \nu}^{\rm YM} \rangle$ in the GZ-theory, it makes sense to calculate the trace anomaly~\cite{Collins:1977} of $\langle T_{\mu \nu}^{\rm YM} \rangle$.  With $m^4 = 2 N g^2 \gamma$, the tree level contibution to the trace anomaly in the GZ-theory is given by
 \beqa
 {\cal A} & = & \left\langle T_{\mu \mu}^{\rm YM} \right\rangle = {4 - d \over 4} \left\langle  (F_{\mu \nu}^b)^2 \right\rangle
 \nonumber \\
 & = & {4 - d \over 4} \left\langle  (\p_\mu A_\nu^b - \p_\nu A_\mu^b)^2 \right\rangle
 \nonumber \\
 & = & \half (N^2 - 1) (4 - d)(d-1) \int {d^d k \over (2 \pi)^d} {(k^2)^2 \over (k^2)^2 + m^4}
 \nonumber \\
 & = &  \half (N^2 - 1) (4 - d)(d-1) \int {d^d k \over (2 \pi)^d} \left( 1 -  {m^4 \over (k^2)^2 + m^4} \right)
 \nonumber \\
& = &  - \half (N^2 - 1) (4 - d)(d-1) \int {d^d k \over (2 \pi)^d} {m^4 \over (k^2)^2 + m^4},
 \eeqa
 where we have subtracted the contribution at the trivial vacuum with $m = 0$\footnote{With dimensional regularization the $m = 0$ term vanishes in any case.}.  We have  
  \beqa
\int {d^d k \over (2 \pi)^d} {m^4 \over (k^2)^2 + m^4} & = & {m^2 \over 2i} \int {d^d k \over (2 \pi)^d} \left( {1 \over k^2 - i m^2} - cc \right)
 \nonumber \\
 & = & {m^2 \over 2i} \int {d^d k \over (2 \pi)^d} \left( \half\int_0^\infty d \alpha \exp[ - \half(k^2 - i m^2) \alpha] - cc \right)
  \nonumber \\
 & = & {m^2 \over 2i} {1 \over (2 \pi)^{d/2}} \left( \half\int_0^\infty d \alpha \alpha^{-d/2} \exp[ - (\epsilon - \half i m^2) \alpha] - cc \right)
  \nonumber \\
 & = & {1 \over (2 \pi)^{d/2}} {m^2 \over i} { \Gamma(3 - d/2) \over (2 - d)(4 - d) } \left[ \left(\epsilon - im^2/2 \right)^{-1 + d/2} - cc \right].
 \eeqa
 The factor  $4-d$ in the denominator cancels the factor $4 - d$ in the coefficient of ${\cal A}$, and in the limit $d \to 4$ of four-dimensional space-time we obtain for the trace anomaly the finite result
 \beq\label{oneloopanom}
 {\cal A} =  - (N^2 - 1) {3 m^4 \over 4 (2 \pi)^2}\ .
 \eeq
 Note that the anomaly comes entirely from $T_{\mu \mu}^{\rm YM}$.  The tree level contribution of the GZ-theory gives the correct sign for the anomaly and implies  that  $\gamma>0$ lowers the vacuum energy density.
 
 We may approximate the physical value of  $m^4 = 2 N g^2 \gamma$ by comparison with QCD sum rule estimates~\cite{Shifman:1979}.  For a pure $SU(N=3)$ gauge theory, the one-loop contribution to the anomaly is,
 \beq\label{anomm}
 {\cal A} = - {3 m^4 \over 2 \pi ^2}.
 \eeq
The expression of the anomaly in terms of the non-perturbative gluon condensate is~\cite{Collins:1977},
\beq
{\cal A} = { \beta(g) \over 2g } \vev{: (F_{\lambda \mu}^a)^2:} \xrightarrow{g^2\rightarrow 0} - {\beta_0 g^2 \over 2 (4\pi)^2 } \vev{ :(F_{\lambda \mu}^a)^2: }
\eeq
where $\beta_0 = 11N/3$. The estimate of the non-perturbative gluon condensate by QCD sum-rules in~\cite{Shifman:1979} implies,
\beqa
{\cal A} & = & - \lim_{g^2\rightarrow 0}{11 g^2 \over 2 (4\pi)^2 } \vev{:(F_{\lambda \mu}^a)^2 :} = - {11 \over 8} {\alpha_s \over \pi} \vev{: (F_{\lambda \mu}^a)^2 :}
\nonumber \\
& \sim & - {11 \over 8} 0.012 GeV^4.
\eeqa
Comparison with \eqref{anomm} then gives the one-loop estimate,
\beq
m^4 = {2 \pi^2  \over 3} {11 \over 8} 0.012 GeV^4,
\eeq
or 
\beq
m = 574 MeV,
\eeq 
a not unreasonable value for the constituent gluon mass. The trace anomaly is a renormalization group invariant (physical) quantity and \eqref{anomm} thus implies that the Gribov parameter is non-perturbative, with $\gamma\sim \pi^2{\cal A}/{3 N g^2}=O(1/g^2)$ at weak coupling.

\section{Kugo-Ojima confinement criterion and BRST}\label{KO-BRST}

 The color charge,  defined in \eqref{QC}, satisfies $[Q_C^a, Q_C^b] = f^{abd} Q_C^d$.  In Faddeev-Popov theory Landau gauge is special in that the color charge is the anti-commutator of the BRST charge, $Q_B$, with another symmetry generator~\cite{Blasi:1991},
\beq\label{comGB}
\{ Q_B, Q_G^a \} = Q_C^a\ .
\eeq
\eqref{comGB} also holds in the GZ-theory with the charge $Q_G^a$  given in \eqref{QG}. $Q_G^a$ shifts the Faddeev-Popov ghost field by a constant, $\{ Q_G^a, c^b(x) \} = \delta^{ab}$.  If $Q_B$ and $Q_G^c$ were well defined, which they are not, and if physical states satisfy $Q_B \ket{ F } = 0$, \eqref{comGB} would prove color confinement, for one has that $ 0 = \vev{ F | \{ Q_B, Q_G^a \} | F } =   \vev{ F | Q_C^a | F }$ for all physical states $\ket{F}$.  The missing parts of this argument are the subject of the  Kugo and Ojima confinement criterion~\cite{Kugo-Ojima, Kugo:1995km}. We here extend, and review for completeness, the analysis in~\cite{Mader:2014} of the Kugo-Ojima confinement criterion~\cite{Kugo-Ojima, Kugo:1995km} and the GZ action.

Consider the gluonic equation of motion, 
\begin{align}
\label{gz_eom1} 
    \var[S]{A_\mu} & = - D_\nu F_{\nu\mu} + i \p_\mu  b +  i\p_\mu {\bar c} \times c      \\ 
      & - \partial_\mu \bar \varphi\times\varphi + \partial_\mu\bar\omega\times\omega 
      - (\partial_\mu \bar\omega)\times\varphi \nonumber\\
      & \ \ \ \ \ \ \   - \gamma^{1/2} f[\bar\omega_\mu]\times c + \gamma^{1/2} f[\vph_\mu-\bar 
      \vph_\mu] \, , \nonumber
\end{align}
where all summed indices have been suppressed and both sides of the equation are in the adjoint representation of global color. Contractions with structure constants in the (upper) color and (lower) flavor indices are denoted by
\be
\cp{\Psi}{\Omega}{a} = g f^{acd} \Psi^{c}_{\mu b} \Omega^{d}_{\mu b} 
\ee
and
\be
\cpt{\Psi}{\Omega}{a} =  g f^{acd} \Psi^{b}_{\mu c} \Omega^{b}_{\mu d} \,.
\ee
The GZ action is invariant under global color transformation $\delta\vartheta^a$ under which $A, c, {\bar c},  b$ transform in the usual way and the auxiliary fields transform in the diagonal subgroup,
\be \delta\Psi^{a}_{\mu b} =  \cp{\Psi_{\mu b}}{\delta\vartheta}{a} +  
\cpt{\Psi_\mu^a}{\delta\vartheta}{a} \label{gz_traf2} \,
\ee
for any 
$\Psi^{a}_{\mu b} \in \{\varphi^{a}_{\mu b},\bar\varphi^{a}_{\mu b},\omega^{a}_{\mu b},
\bar \omega^{a}_{\mu b}\}$.  The corresponding conserved Noether color-current in the adjoint representation is given by
\begin{align}
       j_\mu = & \ A_\lambda \times F_{\lambda \mu} + A_\mu \times i b - i \p_\mu {\bar c} \times c + {i\bar c} \times D_\mu c
      \nonumber\\
	    &  + \partial_\mu \bar \vph\times\vph + \partial_\mu \bar \vph\widetilde\times\vph - 
	    \bar \vph\times D_\mu \vph   \nonumber \\
	    & - \bar \vph\widetilde\times D_\mu \vph - \partial_\mu\bar \omega\times\omega -
	     \partial_\mu\bar \omega\widetilde\times\omega\label{gz_cur} \\
	    &  +  {\bar \omega} \times ( {D_\mu \omega} + D_\mu c \times \vph )  + {\bar \omega} \widetilde{\times} ( {D_\mu \omega} + D_\mu c \times \vph )
	    \nonumber  \\
       & + (\partial_\mu \bar \omega\times\vph)\times c + 
    \gamma^{1/2} f[\bar\omega_\mu]\times c  \,. \nonumber
\end{align}
This allows us to express \eqref{gz_eom1} in terms of the color-current and a BRST exact contribution,
\be \var[S]{A_\mu} = - \partial_\nu F_{\nu\mu} - j_\mu + s\bar\chi_\mu \,, 
\label{gz_eom2}
\ee
where
\beq
\label{definechibar}
 \bar\chi_\mu = iD_\mu \bar c - \bar \omega\times D_\mu \vph  - 
 \bar \omega\widetilde\times D_\mu \vph
  + \partial_\mu \bar \omega\widetilde\times\vph  - \gamma^{1/2} f[\bar\omega_\mu] \,
\eeq
\beqa
\label{defineschibar}
s \bar\chi_\mu & = & i D_\mu b + \gamma^{1/2} f[\varphi_\mu - \bar\varphi_\mu] + i\bar c \times D_\mu c - \bar\varphi \times D_\mu \varphi - \bar\varphi \tilde\times D_\mu \varphi + \p_\mu \bar\varphi \tilde\times \varphi 
\nonumber \\
&&+ \bar\omega \times ( D_\mu \omega + D_\mu c \times \varphi )
+ \bar\omega \widetilde\times ( D_\mu \omega + D_\mu c \times \varphi ) 
- \p_\mu \bar\omega \tilde\times \omega  .
\eeqa
The gluonic quantum equation of motion follows,
\begin{align}
   \delta^{ab}\delta_{\mu\sigma}\delta(x-y) &= \vev{A_{\sigma}^a(y)\,\var[S]{A_{\mu}^b(x)} }
    \label{gz_dse} \nonumber   \\
   & = -\vev{A_{\sigma}^a(y)\,(\partial_\nu F_{\nu\mu}^b + j_\mu^b)(x) }   + \vev{A_{\sigma}^a(y)\,s\bar\chi^b_\mu(x) }\ . 
\end{align}

We could complete the Kugo-Ojima argument for color confinement if the $s$-exact expectation value $\vev{s [A_{\sigma}^a(y)\,\bar\chi^b_\mu(x)] }$ vanishes, as it would if  BRST symmetry is preserved.  Temporarily assuming this is the case, we  rewrite \eqref{gz_dse} in the form,
\begin{align}
   \delta^{ab}\delta_{\mu\sigma}\delta(x-y) &= \vev{A_{\sigma}^a(y)\,\var[S]{A_{\mu}^b(x)} }
    \label{gz_dse2} \nonumber   \\
   & = -\vev{A_{\sigma}^a(y)\,(\partial_\nu F_{\nu\mu}^b + j_\mu^b)(x) }   - \vev{(s A)_{\sigma}^a(y) \, \bar\chi^b_\mu(x) }\ . 
\end{align}
Upon fourier transformation ($\cal FT$), \eqref{gz_dse2} reads,
\beq
\label{saturate}
 \delta^{ab}\delta_{\mu\sigma} =  \delta^{ab} T_{\sigma\mu} f(p^2) +  \delta^{ab}[L_{\sigma\mu}-T_{\sigma\mu}u(p^2) ] -\vev{A_{\sigma}^a\, j_\mu^b }_\F(p)
\eeq
where $L_{\sigma \mu} = p_\sigma p_\mu/p^2$, $T_{\sigma \mu} = \delta_{\sigma \mu} - L_{\sigma \mu}$, and the functions $f(p^2)$ and $u(p^2)$ are defined by
\begin{subequations}
\label{GZformfactors}
\begin{align}
-\vev{A_{\sigma}^a(y) \, \partial_\nu F_{\nu\mu}^b(x) }_\F&= \delta^{ab}T_{\sigma\mu} f(p^2) \label{fGZ}\\
- \vev{(D_\sigma c)^a(y)\,\bar\chi^b_\mu(x)}_\F&= \delta^{ab} [L_{\sigma\mu}-T_{\sigma\mu}u(p^2) ] \label{uGZ}\ .
\end{align}
\end{subequations}
If these functions satisfy,
\be\label{confGZ}
f(0)=0\qquad\text{and}\qquad u(0)=-1\, ,
\ee
the first term in \eqref{saturate} vanishes in the infrared, $\left.\vev{A_{\sigma}^a\,\partial_\nu F_{\nu\mu}^b }_\F\right|_ {p = 0} = 0$, and  the second contribution  saturates \eqref{saturate} in the infrared, $ \left.\vev{(D_\sigma c)^a\,\bar\chi^b_\mu}_\F\right|_{p = 0} = -\delta^{ab} \delta_{\mu \nu}$.  It follows that the matrix element with the color current also vanishes in the infrared, $\left.\vev{A_{\sigma}^a\, j_\mu^b }_\F\right|_{ p = 0} = 0$.  Thus if (\ref{confGZ}) holds, none of these terms has a massless particle pole  and there is no long range color field. Color is confined in a phase where the current matrix element  $\vev{A_\mu(x) j_\mu(y)}$ vanishes for $p^2\rightarrow 0$.  

According to \eqref{comegabarprop} the $c$-$\bar\omega$ propagator at fixed $A$ vanishes.  From \eqref{definechibar} and \eqref{uGZ} we then obtain
\beq
- \vev{(D_\sigma c)^a(y)\,(iD_\mu {\bar c})^b(x)}_\F =- \vev{(D_\sigma c)^a(y)\,\bar\chi^b_\mu(x)}_\F= \delta^{ab} [ L_{\sigma\mu}-T_{\sigma\mu}u(p^2) ] \label{uGZa}\ .
\eeq
Summing \eqref{uGZa} over directions and color we have,  
\beq
\label{KOfunction}
-i\int d^d x \ \exp(-i p \cdot x) \vev{ (D_\mu c)^a(x) (D_\mu {\bar c})^a(0) } = (N^2-1)( (1-d) u(p^2) + 1)\ .
\eeq
The horizon condition of \eqref{horizonconditiona} implies that this expression should equal  $d (N^2 -1)$ at $p^2=0$,  or\footnote{The horizon condition controls the asymptotic behavior of $\vev{ (D_\mu c)^a(x) (D_\nu {\bar c})^b(0) } $ which, according to \eqref{asymptoticbehavior}, falls off at large $x$ more rapidly than $1/|x|^d$.  The canonical dimension of this correlator is $1/|x|^d$, so the horizon condition implies that this correlator is of {\em shorter} range than canonical.  This leads to a ghost propagator $\vev{ c(x) \bar c(y) }$ that is of {\em longer} range than canonical~\cite{Kugo-Ojima, Kugo:1995km}.} that $u(0)= -1$.  

We are not in a position to evaluate the condition $f(0) = 0$ exactly.  However we can evaluate it at tree level, with the result
\beq\label{treeglueon}
p^2 \vev{A_{\sigma}^a(y)\, A_\mu^b (x) }_\F = \delta^{ab}T_{\sigma\mu} {(p^2)^2 \over (p^2)^2 + m^4}, 
\eeq  
where $m^4 = 2 N_c g^2 \gamma$ is the QCD mass related to the tree-level trace anomaly of \eqref{oneloopanom}.  This vanishes at $p = 0$, and thus both conditions of \eqref{confGZ} are satisfied, and with them the Kugo-Ojima confinement criterion.

Quite strikingly, both criteria of \eqref{confGZ} for a confining phase hold already at tree level in the GZ theory. Perturbative calculations to one- and two-loop order in three~\cite{Gracey:2010df} and four~\cite{Ford:2009ar,Gracey:2009zz,Gracey:2009mj} Euclidean dimensions as well as a non-perturbative infrared analysis~\cite{Huber:2009tx} of  the GZ action show that in the infrared the gluon propagator remains suppressed and the ghost propagator diverges more strongly than a massless pole~\cite{Zwanziger:1992qr, Huber:2009tx,Huber:2010}. This infrared behavior agrees with the original Kugo-Ojima scenario~\cite{Kugo:1995km}.  

Till now we have left in abeyance whether BRST is preserved in this instance, that is whether $\vev{s(A_{\sigma}^a\, \bar\chi^b_\mu)} = 0$.  Note that the fields $A$ and $\bar\chi$ are not observables, so our hypothesis that BRST symmetry is unbroken in the physical space is of no avail.  However, we can check by direct calculation whether 
\beq
\label{quartet}
0=\vev{s(A_{\sigma}^a(y)\, \bar\chi^b_\mu(x))} \\
    = \vev{A_{\sigma}^a(y)\, s(\bar\chi^b_\mu(x))}+\vev{sA_\sigma^a(y)\, \bar\chi^b_\mu(x)}
\eeq
holds where it is needed, namely in the infrared limit.  In the infrared the second term is given exactly in \eqref{uGZ} due to the horizon condition,  $\left.\vev{sA_\sigma^a\, \bar\chi^b_\mu}_\F\right|_{p = 0} = -  \delta^{ab} \delta_{\mu \nu}$.  We  may evaluate the longitudinal part of the first term $\vev{\p_\sigma A_{\sigma}^a(y)\, s(\bar\chi^b_\mu(x))}$, using the equation of motion of the $b$ field, with the result $\vev{A_\sigma^a \ s\bar\chi_\mu^b}_{\F}^{\rm longitudinal} = \delta^{ab} p_\sigma p_\mu/p^2$, so the longitudinal part of the identity is satisfied exactly.  We cannot currently evaluate the transverse part of the first term exactly and do so only at tree level.  Only the second term of \eqref{defineschibar} gives a transverse contribution at tree level,\footnote{It contributes at tree level because the trace anomaly of \eqref{oneloopanom} implies that  $g^2 \gamma\sim \Lambda_{\rm QCD}^4$.} $\vev{A_\sigma^a \ s\bar\chi_\mu^b}_{\F }^{\rm transverse} = g \gamma^{1/2} f^{bde} \vev{ A_\sigma^a(y) (\varphi - \bar\varphi)^d_{\mu e}(x) }$. With the propagator,
\beq
\vev{ A_\sigma^a(x) (\varphi^b_{\mu c} - \bar\varphi^b_{\mu c})(y) }_\F  =  2 g \gamma^{1/2} f^{abc} {T_{\sigma \mu} \over (p^2)^2 + m^4 },
\eeq
the transverse part of the first correlator in \eqref{quartet} at tree-level is,
\beq
\vev{A_\sigma^a \ s\bar\chi_\mu^b}_{\F }^{\rm transverse} = \delta^{ab} T_{\sigma \mu} {m^4 \over (p^2)^2 + m^4} \xrightarrow{p = 0} \delta^{ab} T_{\sigma \mu}\ .
\eeq
It follows that to leading order,
\beq
\left.\vev{s[A_\mu^a(x)  \ \bar\chi^b_\nu(y)]}_{\F}\right|_{p \rightarrow 0} = 0.
\eeq
This is another instance where the BRST symmetry of the GZ theory appears to be preserved where needed. Although the hypothesis of \eqref{hypothesis} is of little use because $s[A^a_\mu(x) \bar\chi^b_\nu(y)]\notin\Wphys$, this s-exact correlator apparently vanishes in the infrared and implies color confinement in the GZ-theory {\'a} la Kugo and Ojima. 


\section{Conclusion}\label{conclusion}

BRST symmetry plays a central role in continuum gauge theory.  It is used to define the physical space, to derive physical Ward identities like that satisfied by the energy-momentum tensor, to obtain the Kugo-Ojima confinement criterion, etc.  These familiar features of standard QCD are jeopardized in the GZ model by the non-vanishing vacuum expectation value of some $s$-exact operators, $\vev{ sX } \neq 0$, which implies that BRST symmetry is spontaneously broken. We here addressed this problem.

Our starting point was the analysis of spontaneous breaking of BRST in the GZ theory.  To better define the theory, the GZ action was quantized on a finite volume with periodic boundary conditions that break the BRST symmetry explicitly.
 In \eqref{sLbreak} the BRST-breaking was expressed as an integral over the surface of the quantization volume of a certain correlator.  From this expression we derived in \eqref{sufficient} a sufficient condition for the expectation value of a BRST-exact functional to vanish when the boundary recedes to infinity.

The GZ model exhibits a large class of unphysical symmetries that act on its ghost fields only.  All physical observables are invariant under these symmetries of the ghost fields as well as the BRST-symmetry.  This sharpens the notion of an observable in the GZ-theory. The definition of the space of observables $\Wphys$ given in \eqref{physicalops} relegates to the unphysical sector of the theory all cases of BRST symmetry breaking we examined. As a working hypothesis we therefore propose that BRST symmetry is preserved in the space $\Wphys$ of observables, that is, $\vev{s X} = 0$, for all $s$-exact operators $sX \in \Wphys$.  This hypothesis  was found to be sufficient for reconstructing the physical Hilbert space from the  observable correlators.

We derived the energy-momentum tensor $T_{\mu \nu} = T_{\mu \nu}^{\rm YM} + s \Xi_{\mu \nu}$ of the GZ theory, and verified the hypothesis for this case  by proving that its BRST-exact part is an observable with vanishing expectation value, $\vev{s \Xi_{\mu \nu}}=0 $.  The horizon condition was essential for this result.  The tree level contribution to the trace anomaly $\vev{T_{\mu \mu}} = \vev{T_{\mu \mu}^{\rm YM}}$ subsequently obtained in \eqref{oneloopanom} is finite and provides a reasonable estimate of the Gribov mass parameter. The sign of the trace anomaly indicates that the vacuum with a positive Gribov mass has lower energy density. In contrast to Faddeev-Popov theory, the GZ-theory satisfies the Kugo-Ojima criteria for color confinement already at tree level.

Several questions remain.  One would like to prove the hypothesis that BRST is preserved by the physical observables.  Though not strictly necessary, one also  might wish to identify a larger class of $s$-exact operators whose expectation value vanishes.  One such instance is found in Appendix~\ref{applycriterion}, another in the derivation the Kugo-Ojima confinement criterion. One would like to know if BRST is preserved in other, similar instances.  Currently this is not easy to verify,  because BRST was generally found to be preserved by certain functionals only when the non-perturbative horizon condition is satisfied exactly.  The question of reflection positivity needed to establish non-perturbative unitarity also has not been addressed by the present article.    

We would like to point out certain parallels in the construction of the Faddeev-Popov and the GZ theories.  The Faddeev-Popov ghosts are introduced to localize the otherwise non-local Faddeev-Popov determinant.  They not only bring new, unphysical, degrees of freedom into the theory, but also a new symmetry, the BRST symmetry.  The unphysical degrees of freedom of the Faddeev-Popov ghosts are excluded from the physical space by requiring that observables be BRST-invariant.  The auxiliary ghosts of the GZ model were introduced to similarly localize the non-local cut-off at the Gribov horizon.  Like the Faddeev-Popov ghosts, they bring new unphysical degrees of freedom into the theory, but also new symmetries, the ghost symmetries. The new, unphysical degrees of freedom of the auxiliary ghosts are excluded from the physical space by requiring observables to be invariant not only under the BRST symmetry, but also under all ghost symmetries.

Although many questions remain, we are impressed by the consistency of the present construction of the physical state space of GZ theory and of the results obtained.  It is particularly intriguing that the BRST symmetry is preserved in cases of physical interest only if the horizon condition is satisfied.  Our results suggest that the spontaneous breaking of the BRST symmetry may be relegated to the unphysical sector, and that the BRST-symmetry preserving physical sector may provide a consistent non-perturbative quantization of gauge theories.

{\vspace{2em}\noindent\bf Acknowlegements: }MS would like to thank members of the Physics Department of New York University for their hospitality.  We are very grateful to the anonymous referee for pointing out a significant omission in the original manuscript.

\appendix

\section{Phantom Symmetries and their Generators}
\label{phantomsymm}

By definition, \eqref{physicalops} or (\ref{physicalopsa}), phantom symmetries are unobservable symmetries of the theory and every observable commutes with all phantom charges. Phantom symmetries nevertheless are a crucial part of the theory. They provide the framework and govern the dynamics and structure of the local quantum field theory,  but, like the trusses of some buildings, remain invisible.

We denote charges and generators of phantom symmetries by $Q_X$, with the subscript $X$ identifying the symmetry. Here we collect and give a short overview over the complete set of phantom symmetries of the GZ-theory\footnote{For $\gamma=0$ some of these were previously used in~\cite{Schaden:1994} to prove renormalizability of the GZ-theory.}. 

The most prominent and important of the phantom symmetries is the BRST symmetry. It is a nilpotent symmetry generated by the fermionic charge $Q_B$ given in \eqref{QB}. All observables are expected to be BRST-invariant. Although it governs the structure and renormalizability of the theory, the BRST symmetry fundamentally is an unobservable symmetry of a gauge theory and thus its most prominent phantom (symmetry).

Nilpotency means that the BRST charge anticommutes with itself,
\beq\label{nilpotency}
\{Q_B,Q_B\}=0
\eeq
as can be verified using \eqref{QB}. Together with the ghost number, $Q_{\cal N}$, this symmetry forms a BRST-doublet (see \eqref{ghostnumber} below). All other phantom symmetries are BRST-doublets as well.  We also classify phantom symmetries by the irreducible representation of the rigid color group generated by the charges $Q^a_C$ defined in \eqref{QC} below.

\subsection{Color-singlet phantom symmetries}
\label{Singletsymm}
Generators of color-singlet phantom symmetries commute with all color charges $Q^a_C$ of \eqref{QC}. The first of these is the BRST symmetry. The ghost number extended to include the auxiliary ghosts is another,
\beq
\label{ghostnumber}
Q_{\mathcal{N}}\equiv \int d^dx \left[ c\cdot {\delta  \over \delta c}-\bar c \cdot {\delta  \over \delta \bar c}+\omega_{\mu}\cdot {\delta  \over \delta\omega_{\mu} }-\bar\omega_{\mu}\cdot {\delta  \over \delta\bar\omega_{\mu}}\right] \ .
\eeq 
Since $[Q_\mathcal{N},Q_B]=Q_B$, these two symmetries  form a BRST doublet $(Q_\mathcal{N},Q_B)$. 
 
In Faddeev-Popov theory in Landau gauge, BRST and anti-BRST and an SL(2,R) symmetry that includes the ghost number as one of its generators \cite{Ojima:1980,Delbourgo:1981,Baulieu:1981} exhaust the color singlet phantom symmetries.  Although the GZ-model implements Landau gauge, this theory remarkably does not appear to possess an anti-BRST symmetry. The analog of the SL(2,R) generator with ghost number -2 of Faddeev-Popov theory is similarly missing.  However, the charge 
\beq
\label{Qplus}
Q_+\equiv \{Q_B,Q_\text{cb}\}= \int d^dx\; c\cdot {\delta  \over \delta \bar c}-\half(c\times c)\cdot {\delta  \over \delta b}\ ,
\eeq
with 
\beq\label{Qcb}
Q_\text{cb}\equiv\int d^dx\; c\cdot{\delta  \over \delta b}\ ,
\eeq
--- which in FP-theory is part of an SL(2,R) multiplet --- is readily verified to also generate a symmety of the GZ-action that changes the ghost number by two.  $(Q_\text{cb},Q_+)$ is a BRST-doublet. Interestingly and contrary to all other BRST-doublets, $Q_\text{cb}$ is an on-shell symmetry of the GZ-action that holds only for transverse gauge field configurations. 

The model in addition exhibits a number of color-singlet phantom symmetries without analog in Faddeev-Popov theory. One of these counts the net number of  auxiliary fields,
\begin{align}
\label{auxnumber}
Q_{\text{aux}}\equiv \int d^dx &\left[ \bar\varphi_\mu\cdot {\delta  \over \delta \bar\varphi_\mu}-\varphi_\mu\cdot {\delta  \over \delta \varphi_\mu}+\bar\omega_{\mu}\cdot {\delta  \over \delta\bar\omega_{\mu}}-\omega_{\mu}\cdot {\delta  \over \delta\omega_{\mu}}\right.\nonumber\\
&\hspace{2em}\left.+\gamma^{1/2}x_\mu\text{Tr}\Big[{\delta  \over \delta \bar\varphi_\mu}+{\delta  \over \delta \varphi_\mu} + i  \bar\varphi_\mu\times   {\delta  \over \delta b}  + i {\bar\omega}_\mu \times  {\delta  \over \delta {\bar c}}\Big]\right] \ .
\end{align}
The terms proportional to $\gamma^{1/2}$ arise from the shift of \eqref{MS} (also see \eqref{shiftpartials}). The charge $Q_{\text{aux}}$ is BRST-exact,
\beq\label{exactaux}
Q_{\text{aux}}=\{Q_B,Q_T\}\ ,
\eeq
with
\begin{align}
\label{QT}
Q_{T}\equiv \int d^dx &\left[ \bar\omega_\mu\cdot {\delta  \over \delta \bar\varphi_\mu}-\varphi_\mu\cdot {\delta  \over \delta \omega_\mu}+\gamma^{1/2}x_\mu\text{Tr}\Big[{\delta  \over \delta\omega_\mu} + i  \bar\omega_\mu\times   {\delta  \over \delta b} \Big]\right] \ ,
\end{align}
and $(Q_{T}, Q_{\text{aux}})$ is a BRST-doublet of scalar color singlet charges. Note that $Q_T$ is a nil-potent scalar charge with ghost number $-1$, but doesn't (anti-) commute with $Q_B$ and thus does not qualify as an anti-BRST generator. However,  it's cohomology does not include auxiliary fields~\cite{Dixon:1977,Barnich:2000zw}, which are simple doublets under the grading of $Q_\text{aux}$. Reconstruction of the physical space of the theory thus could be viewed as an equivariant one: on the cohomology of $Q_T$,  $Q_B$ is unbroken and equivalent to the BRST-charge of Faddeev-Popov theory. Its cohomology in the sector with vanishing ghost number thus are the gauge-invariant functionals only.

The invariance of the Lagrangian density, \eqref{SGZ}, with respect to the variation $\delta\bar\varphi^a_{\mu b}=\bar\epsilon_\mu \delta^a_b$ with $\p_\nu\bar\epsilon_\mu=0$ is another new phantom symmetry. This shift by a constant color-singlet vector is generated by,
\beq\label{Qphibar}
Q_{\bar\varphi,\mu} \equiv \int d^dx \ \text{Tr} {\delta \over \delta \bar\varphi_\mu} .
\eeq

The commutator of $Q_{\bar\varphi, \mu}$  with the BRST charge $Q_B$ of \eqref{QB} gives an associated phantom symmetry generated by,  
\beq
\label{Qwbar}
Q_{\bar\omega,\mu}=-s Q_{\bar\varphi,\mu}=-[Q_B,Q_{\bar\varphi,\mu}] \equiv \int d^dx \ \text{Tr}{\delta \over \delta \bar\omega_\mu} ,
\eeq
and $(Q_{\bar\varphi,\mu},Q_{\bar\omega,\mu})$ are a BRST-doublet of color-singlet charges. Note that invariance of observables under $Q_B$ \emph{and} $Q_{\bar\omega,\mu}$ implies their invariance under $Q_B+ a_\mu Q_{\bar\omega,\mu}$ for an arbitrary constant vector $a_\mu$. The choice of origin in the definition of the BRST-charge $Q_B$ in \eqref{QB} thus is of no import for operators that commute with the phantom symmetry generator~$Q_{\bar\omega,\mu}$.

At transverse configurations,  $\p_\mu A_\mu=0$, the GZ-action of \eqref{SGZ} by inspection  also is  invariant with respect to the shift   $\delta \omega^a_{\mu b}=\epsilon_\mu \delta^a_b$. This color singlet phantom symmetry is generated by the charge,
\beq\label{Qw}
Q_{\omega, \mu} \equiv \int d^dx \ {\mathfrak{q}}_\mu,
\eeq
with the density,
\beq\label{tq}
\mathfrak{q}_\mu \equiv  \text{Tr}\Big[ {\delta \over \delta {\omega}_\mu}  + i {\bar\omega}_\mu\times{\delta \over \delta b}\Big]\ . 
\eeq
The BRST-variation of $Q_{\omega, \mu}$ is,
\beq\label{Qphi}
Q_{\varphi, \mu} =s Q_{\omega, \mu}=\{Q_B,Q_{\omega, \mu}\}\equiv \int d^dx\ \mathfrak{p}_\mu 
\eeq
with the density,
\beq
\label{tp}
\mathfrak{p}_\mu = s\mathfrak{q}_\mu= \text{Tr}\Big[{\delta  \over \delta \varphi_\mu} + i  \bar\varphi_\mu\times   {\delta  \over \delta b}  + i {\bar\omega}_\mu \times  {\delta  \over \delta {\bar c}}\Big]\ .
\eeq
$(Q_{\omega, \mu},Q_{\varphi, \mu})$ thus is another BRST-doublet of color-singlet phantom generators.

Finally, at $\gamma=0$,  ${\cal L}^{\rm gf}$ evidently is invariant under the internal symmetry $\delta \omega^a_{\mu b}=\epsilon_{\mu\nu} \varphi^a_{\nu b}; \delta \bar\varphi^a_{\mu b}=\epsilon_{\mu\nu} \bar\omega^a_{\nu b}$ with antisymmetric  $\epsilon_{\mu\nu}=-\epsilon_{\nu\mu}$ that rotates anti-commuting and commuting auxiliary fields into each other. This phantom (super-)symmetry persists for  $\gamma> 0$ in an extended form generated by,\footnote{The symmetries of the theory are most easily found in terms of the unshifted variables by inspection of the unshifted Lagrangian, \eqref{totallagrangian} and (\ref{PsiB}), where they are manifest ($\gamma = 0$).  They may then be expressed in terms of the shifted variables.}
\beq
\label{QN}
Q_{N, \mu\nu}\equiv  
\int d^dx \ \left[ \varphi_{\nu}\cdot {\delta  \over \delta \omega_{\mu}}
+ {\bar\omega}_{\nu} \cdot{\delta  \over \delta \bar\varphi_{\mu}} 
 + \gamma^{1/2} x_\mu \mathfrak{q}_\nu \right] - [\mu \leftrightarrow \nu].
\eeq
The BRST-variation of $Q_{N, \mu\nu}$ gives the generators of an internal $SO(4)$ symmetry of the auxiliary ghost, 
\beq\label{QM}
Q_{M, \mu \nu} =\{Q_B, Q_{N, \mu\nu}\} \equiv  \int d^dx \ \left[  \mathfrak{s}_{\mu \nu} +\gamma^{1/2} x_\mu (\mathfrak{p}_\nu-\text{Tr} {\delta  \over \delta \bar\varphi_\nu})\right] -[\mu\leftrightarrow\nu]
\eeq
with densities
\beq
\label{sdensity}
\mathfrak{s}_{\mu \nu} \equiv \varphi_{\nu}\cdot {\delta  \over \delta \varphi_{\mu}}
+ \bar\varphi_{\nu}\cdot  {\delta  \over \delta \bar\varphi_{\mu}} 
 + \omega_{\nu}\cdot  {\delta  \over \delta \omega_{\mu}} 
      + {\bar\omega}_{\nu}\cdot  {\delta  \over \delta {\bar\omega}_{\mu}} \ .
\eeq
and $\mathfrak{p}_\mu$ given in \eqref{tp}.

These color singlet phantom symmetry generators form a closed algebra. They appear to exhaust the color-singlet phantom symmetries of the GZ-theory.  Like $Q_B$, many are spontaneously broken, for example 
\beq\label{brokenQwbar}
\langle [Q_{\bar\omega, \nu}, {\bar\omega}^a_{\mu b}] \rangle = \delta_{\mu \nu} \delta^a_b.
\eeq

\subsection{Color-adjoint phantom symmetries}
\label{coloradjointphantoms}
Faddeev-Popov theory in Landau gauge and GZ-theory share the remarkable property~\cite{Blasi:1991}, that the color charge $Q_C$ is BRST-exact. With the BRST-charge of \eqref{QB} we have that~\cite{Schaden:1994}, 
\beq
\label{QCexact}
Q_C^a =   \{Q_B, Q_G^a\} \ ,
\eeq
where
\beqa\label{QG}
\hspace{-2em}Q_G \equiv \int d^dx \ \left[  -  {\delta \over \delta c} +  {\bar c} \times  {\delta \over \delta  b}  +  \varphi_{\mu a}  \times  {\delta \over \delta \omega_{\mu a}} +  {\bar\omega}_{\mu a}\times {\delta \over \delta \bar\varphi_{\mu a}}   + \varphi^a_\mu \widetilde\times {\delta \over \delta \omega^a_\mu}  + {\bar\omega}^a_\mu \widetilde\times  {\delta \over \delta \bar\varphi^a_\mu}  \right]\ ,
\eeqa
and  the generator of rigid color rotations, $Q_C$, of the GZ-theory is ~\cite{Schaden:1994,Mader:2014},
\beqa
\label{QC}
Q_C& \equiv & \int d^dx \Big[ A_\mu \times  {\delta \over \delta A_\mu} + c \times  {\delta \over \delta c} + \bar c \times  {\delta \over \delta \bar c} + b \times  {\delta \over \delta b}
+ \varphi_{\mu a} \times  {\delta \over \delta \varphi_{\mu a}} + \bar\varphi_{\mu a} \times  {\delta \over \delta \bar\varphi_{\mu a}}\nonumber\\
&&\hspace{-2em} + \omega_{\mu a} \times  {\delta \over \delta \omega_{\mu a}} + \bar\omega_{\mu a} \times  {\delta \over \delta \bar\omega_{\mu a}}
 + \varphi^a_\mu \widetilde\times  {\delta \over \delta \varphi^a_\mu} + \bar\varphi^a_\mu \widetilde\times  {\delta \over \delta \bar\varphi^a_\mu} + \omega^a_\mu \widetilde\times  {\delta \over \delta \omega^a_\mu} + \bar\omega^a_\mu \widetilde\times  {\delta \over \delta \bar\omega^a_\mu} \Big]\ .
\eeqa

These generators of an unbroken diagonal $SU(N)$-symmetry are the sum of the charges, $Q_K$, of the $SU(N)$ symmetry of rigid color rotations and of the $SU(N)$ subgroup of flavor symmetry generators given in \eqref{QF},
\beq\label{QCdecom}
Q_C^a=Q_K^a+ \half g f^{abc}Q_{F,\mu\mu bc}\ , 
\eeq
with 
\beqa\label{QK}
Q^a_K&\equiv& \int\! d^dx \Big[\Big(A_\mu \!\times\!   {\delta \over \delta A_\mu} \!+\!  c \!\times\!   {\delta \over \delta c} \!+\!  \bar c \!\times\!   {\delta \over \delta \bar c} \!+\!  b \!\times\!   {\delta \over \delta b}
\!+\!  \varphi_{\mu b} \!\times\!   {\delta \over \delta \varphi_{\mu b}} \!+\!  \bar\varphi_{\mu b} \!\times\!   {\delta \over \delta \bar\varphi_{\mu b}} \!+\!  \omega_{\mu b} \!\times\!   {\delta \over \delta \omega_{\mu b}}\nonumber\\
&&\hspace{-1em} \!+\!  \bar\omega_{\mu b} \!\times\!   {\delta \over \delta \bar\omega_{\mu b}}\Big)^a
-g \gamma^{1/2}x_\mu f^{abc}\left[\Big(i \bar\varphi_{\mu c}\!\times\! {\delta  \over \delta b}\!+\! i \bar\omega_{\mu c}\!\times\! {\delta  \over \delta \bar c}\Big)^b\!+\!  {\delta  \over \delta \bar\varphi^b_{\mu c}}\!+\! {\delta  \over \delta \varphi^b_{\mu c}}\right]-i g^2\gamma N x^2 {\delta  \over \delta b^a}\Big]\ .\nonumber\\
\eeqa
Note that the symmetry of rigid color rotations, generated by $Q_K$, by itself is spontaneously broken, whereas the diagonal group that includes flavor rotations generated by $Q_C$ remains unbroken. 

As for $Q_C$, the charges $Q_G$ may be decomposed, 
\beq\label{QGdecom}
Q_G^a=Q_J^a+ \half g f^{abc}Q_{E,\mu\mu bc}\ , 
\eeq
with
\beq\label{QJ}
Q_J \equiv \int d^dx \ \Big[  -  {\delta \over \delta c} +  {\bar c} \times  {\delta \over \delta  b}  +  \varphi_{\mu a}  \times  {\delta \over \delta \omega_{\mu a}} +  {\bar\omega}_{\mu a}\times {\delta \over \delta \bar\varphi_{\mu a}}  -g\gamma^{1/2}f^{abc} x_\mu\left( {\delta  \over \delta\omega_{\mu c}}+i  \bar\omega_{\mu c}\times {\delta  \over \delta b}\right)^{\!\! b}\,\Big]\ .
\eeq
 The generators $Q_K$ and $Q_J$ for $\gamma=0$ were used in~\cite{Schaden:1994} to show renormalizability of the theory.    

The nilpotent BRST-operator $Q_B$ thus commutes with the generators $Q_C^a$, and the pairs $(Q_G^a,Q_C^a)$ and $(Q^a_J,Q^a_K)$ are BRST doublets. The phantom charges of \eqref{QC} and \eqref{QG} form adjoint multiplets,
\beq\label{adjghost}
[Q^a_C,Q^b_C]=f^{abc} Q^c_C\ , \ \ [Q^a_C,Q^b_G]=f^{abc} Q_G^c\ , \ \ [Q^a_G,Q^b_G]=0\ .
\eeq

The GZ-theory in addition possesses a set of phantom symmetries in the adjoint that mix auxiliary vector ghosts with  FP-ghosts. Inspection of ${\cal L}^{\rm gf}$ in \eqref{SGZ} reveals the invariance, $\delta \bar c^a=i\eps_{\mu b} \bar\omega^a_{\mu b}, \delta\omega^a_{\mu b}=\eps_{\mu b} c^a$. This phantom symmetry is generated by the charges, 
\beq\label{QR}
Q_{R,\mu a} \equiv\int d^dx [c\cdot {\delta  \over \delta \omega_{\mu a}}
+i\bar\omega_{\mu a}\cdot{\delta  \over \delta \bar c}]\ .
\eeq
of vanishing ghost number.  Commuting with the BRST-generator $Q_B$ reveals another set of phantom generators in the adjoint representation,
\beq\label{QS}
Q_{S,\mu a} \equiv -[Q_B,Q_{R,\mu a}]= \int d^dx [\half (c\times c)\cdot {\delta  \over \delta \omega_{\mu a}}+c\cdot {\delta  \over \delta \varphi_{\mu a}}
-i\bar\varphi_{\mu a}\cdot {\delta  \over \delta \bar c}-i\gamma^{1/2} x_\mu {\delta  \over \delta \bar c^a}]\ ,
\eeq
which are readily verified to commute with the GZ-action of  \eqref{SGZ}.

$Q_{R,\mu a}$, defined in \eqref{QR}, is one of the more interesting unbroken phantom symmetries. It is due to the absence of a  vertex containing $\bar c$ and $\omega$ in ${\cal L}^{\rm gf}$ and implies that any operator with a positive number of FP-ghosts $\#c-\#\bar c$ has vanishing expectation value.  This is true at fixed $A$, after integrating out the ghost fields, as is explained following \eqref{comegabarprop}. In particular the $\vev{c\bar\omega}_A$-propagator for a fixed $A$-field vanishes in the GZ theory.  We believe that $Q_C, Q_N$ and $Q_{R, \mu a}$ are the only phantom symmetries that are not broken spontaneously.

Another doublet of adjoint phantom symmetries is found by (anti-)commuting these generators with color singlet phantoms. This closes the subalgebra with the additional commutation relations, 
\beq\label{AdSing}
\{Q_{\bar\omega,\mu},Q_{R,\nu}\}=-[Q_{\bar\varphi,\mu},Q_{S,\nu}]= i\delta_{\mu\nu} Q_{\bar c}=-i\delta_{\mu\nu} [Q_B,Q_{NL}]=\delta_{\mu\nu}[Q_+,Q_G]\ ,
\eeq
and the adjoint phantom charges,
\beq\label{QbQcbar}
Q^a_{NL}\equiv\int d^dx {\delta  \over \delta b^a}\ \ \ \text{and }\ \ Q^a_{\bar c}\equiv\int d^dx{\delta  \over \delta \bar c^a}\ .
\eeq
It is worth noting in this context that $Q_G^a,Q_{NL}^a$ as well as $Q^a_{\bar c}$ generate broken symmetries even in Faddeev-Popov theory in Landau gauge, since
\beq\label{brokenFP}
\langle \{Q^a_G, c^b(x)\}\rangle= \langle \{Q^a_{\bar c}, \bar c^b(x)\}\rangle= \langle \{Q^a_{NL}, b^b(x)\}\rangle=\delta^{ab}\neq 0\ .
\eeq
The corresponding Goldstone zero-modes do not couple to observables and are often ignored (see, however,~\cite{Sharpe:1984,Baulieu:1998, Zwanziger:2010a}). That $Q_{\bar\omega,\nu}$ is the charge of a broken symmetry due to \eqref{brokenQwbar}, in this sense is not extraordinary.

For $SU(N>2)$ there are four more adjoint multiplets of BRST-doublets, $(Q_{E_A},Q_{F_A})$, $(Q_{E_S}, Q_{F_S})$, $(Q_{U_A},Q_{V_A})$, $(Q_{U_S}, Q_{V_S})$ using the symmetric combinations.

\subsection{Additional phantom symmetries}  

At $\gamma=0$ the GZ-theory is invariant under an internal $U(d(N^2-1))$ symmetry of the auxiliary fields that acts on pairs $B=(\mu, b)$ of ``vector-flavor" indices and transforms fermionic into bosonic ghosts and vice versa.  For $\gamma>0$ these symmetries are generated by,
\beq\label{QE}
Q_{E,\mu\nu\, ab}\equiv \int d^dx \left[\bar\omega_{\nu b}\cdot{\delta  \over \delta \bar\varphi_{\mu a}}-\varphi_{\mu a}\cdot{\delta  \over \delta \omega_{\nu b}}+\gamma^{1/2} x_\mu\left( {\delta  \over \delta \omega_{\nu b}}+i  \bar\omega_{\nu b}\times {\delta  \over \delta b}\right)^{\!\! a}\,\right]\ .
\eeq
 The BRST-variation of these charges is,
\beqa\label{QF}
Q_{F,\mu\nu\, ab}&\equiv&\{Q_B,Q_{E,\mu\nu\, ab}\}= \int d^dx \Big[\bar\varphi_{\nu b}\cdot{\delta  \over \delta \bar\varphi_{\mu a}}-\varphi_{\mu a}\cdot{\delta  \over \delta \varphi_{\nu b}}+\bar\omega_{\nu b}\cdot{\delta  \over \delta \bar\omega_{\mu a}}-\omega_{\mu a}\cdot{\delta  \over \delta \omega_{\nu b}}\nonumber\\
&&\hspace{-5em}+\gamma^{1/2} x_\nu{\delta  \over \delta \bar\varphi^b_{\mu a}}+\gamma^{1/2} x_\mu\left({\delta  \over \delta \varphi_{\nu b}}+i  \bar\varphi_{\nu b}\times{\delta  \over \delta b}+i \bar\omega_{\nu b}\times{\delta  \over \delta \bar c}\right)^a+i \gamma x_\mu x_\nu g f^{abc} {\delta  \over \delta b^c}\Big]\ .
\eeqa
An additional BRST-doublet  is revealed by translating $x_\mu\rightarrow x_\mu+a_\mu$ in \eqref{QE} and \eqref{QF} by an arbitrary constant vector,
\beqa\label{QUQV}
Q^a_{U,\nu b}\equiv[Q_{R,\nu b}\,,Q^a_G]&=&\int d^dx \Big({\delta  \over \delta \omega_{\nu b}}+i \bar\omega_{\nu b}\times {\delta  \over \delta b}\Big)^{\!\! a}\\ Q^a_{V,\nu b}\equiv \{Q_B,Q^a_{U,\nu b}\}&=& \int d^dx \left[\Big({\delta  \over \delta \varphi_{\nu b}}+i\bar\varphi_{\nu b}\times{\delta  \over \delta b}+i \bar\omega_{\nu b}\times{\delta  \over \delta \bar c}\Big)^{\!\! a}+i \gamma^{1/2}x_\nu g f^{abc} {\delta  \over \delta b^c}\right]\nonumber\ .
\eeqa
Note that these charges are needed to close the algebra.

Some of the color-singlet components of these phantom charges we already encountered in \eqref{auxnumber}, \eqref{QT}, \eqref{Qw}, \eqref{Qphi}, \eqref{QN}, \eqref{QM}, \eqref{Qw}, \eqref{Qphi} and \eqref{Qphibar}. There are additional \emph{symmetric} and traceless color singlet charges which we do not separately enumerate here. 

The structure constants $f^{abc}$ and $d^{abc}$, project on the additional phantom charges in the adjoint representation mentioned at the end of Sec.~\ref{coloradjointphantoms}, 
\begin{align}\label{adjoints}
Q^a_{{E_A},\mu\nu}&=f^{abc}Q_{E,\mu\nu\, bc} &  Q^a_{{E_S},\mu\nu}&=d^{abc}Q_{E,\mu\nu\, bc}\nonumber\\
Q^a_{{F_A},\mu\nu}&=f^{abc}Q_{F,\mu\nu\, bc}  &  Q^a_{{F_S},\mu\nu}&=d^{abc}Q_{F,\mu\nu\, bc}\nonumber\\
Q^a_{{U_A},\mu}&=f^{abc}Q^b_{U,\mu c}  &  Q^a_{{U_S},\mu}&=d^{abc}Q^b_{U,\mu c}\nonumber\\
Q^a_{{V_A},\mu}&=f^{abc}Q^b_{V,\mu c} &  Q^a_{{V_S},\mu}&=d^{abc}Q^b_{V,\mu c}\ .
\end{align}
For an $SU(3)$ gauge theory one further can project $(Q_E,Q_F)$ and $(Q_U,Q_V)$ onto two $\mathbf{10}$ and one $\mathbf{27}$ multiplets. We here refrain from doing so because these higher irreducible representations of phantom symmetries depend on the gauge group and we have no explicit use for them. 

\subsection{Some comments on phantom symmetries}
The set $\mathfrak{F}$ of $2d+3 +2(d+2)(N^2-1)+2d(d+1)(N^2-1)^2$ linearly independent generators,
\beq\label{Fset}
\mathfrak{F}=\{Q_B,Q_\mathcal{N},Q_+, Q_{\bar\varphi,\mu},Q_{\bar\omega,\mu},Q^a_G,Q^a_C,Q^a_{NL},Q^a_{\bar c},Q_{R,\mu a},Q_{S,\mu a},Q_{E,\mu\nu\, ab},Q_{F,\mu\nu\, ab},Q^a_{U,\mu b},Q^a_{V,\mu b}\}
\eeq
forms a closed algebra of unphysical charges. This set seems to exhausts all phantom symmetry generators of the GZ-theory
For $SU(3)$ and 4-dimensional space-time, the model thus has an algebra of $2667$ phantom charges. Faddeev-Popov theory in Landau gauge by contrast has a mere $37$ phantom generators (BRST, anti-BRST, the SL(2,R), as well as ghost number and the analogs of $Q^a_G,Q^a_C,Q^a_{\bar c}$ and $Q^a_{NL}$). But the GZ-model also includes $4 d (N^2-1)^2$ or $1024$ more unphysical fields than Faddeev-Popov theory. 

If we consider $(Q_\mathcal{N},Q_B)$ a doublet, all phantom generators come in BRST doublets\footnote{Although $Q_\text{cb}$ of \eqref{Qcb} does not generate a symmetry of the action, $(Q_\text{cb}, Q_+)$ algebraically are a BRST-doublet.} .  Phantom symmetries therefore should be unphysical and unobservable~\cite{Dixon:1977,Barnich:2000zw}. Remarkably, the color charge of the model is part of a BRST doublet, yet another indication that color may not be observable~\cite{Kugo-Ojima, Mader:2014}.

In general it is a daunting algebraic task to determine whether an operator $F$ is invariant under all the phantom symmetries of \eqref{Fset} and belongs to the set ${\cal W}_{\rm phys}$ defined in \eqref{physicalops}. Verifying the $U(d(N^2-1))$ symmetry that rotates the auxiliary ghosts into each other can be helpful in this regard. 

That phantom symmetries are BRST-doublets can also be exploited. The Jacobi identity implies invariance under the generator $[Q_X, Q_Y]$ if an  operator is invariant under the symmetries $Q_X$ and $Q_Y$ separately. To establish whether an operator $F$ is in $\Wphys$ one thus need only show that $F$ commutes with the set of charges,
\be\label{smallF}
\widetilde{\mathfrak{F}}=\{Q_+,Q_B, Q_{\cal N},Q_{\bar\varphi,\mu},Q^a_G,Q^a_{NL},Q_{R,\mu a},Q_{E,\mu\nu\, ab}\}\ ,
\ee
whose (graded) commutators generate the whole algebra of phantom charges. Note that the subsets $\{Q_+\},\{Q_B\},\,\{Q_{\cal N}, Q_{\bar\varphi,\mu},Q^a_{NL},Q_{R,\mu a}\},\,\{Q^a_G,Q_{E,\mu\nu\, ab}\}$ of $\widetilde{\mathfrak{F}}$, with ghost number $2,1,0,-1$ respectively, are sets of mutually (anti-)commuting charges.

All the phantom symmetries associated with charges in $\widetilde{\mathfrak{F}}$ except (possibly) those generated by the ghost number $Q_{\cal N}$, global color $Q_C^a$, $Q_+$ and by $Q_{R,\mu a}$, are spontaneously broken in the GZ-theory. However, like the phantom symmetries themselves, this breaking is hidden and the corresponding Goldstone excitations are unobservable.

\section{Check of the surface equation}\label{checksurfaceeq}

Let us verify that eq.~(\ref{sLbreak}) holds for the case $F = \bar\omega^a_{\mu b}(y)$.  We must verify
\beq
\label{sLiscacul}
\vev{s \bar\omega^a_{\mu b}(y) } =\gamma^{1/2} \frac{L}{2}\sum_{\mu =1}^d\sum_{\sigma=\pm} \int_{x_\mu=\sigma L/2}\hspace{-2em} dS_\mu \  \vev{ \Tr (D_\nu \omega_\nu +  D_\nu c \times \varphi_\nu)(x) \ \bar\omega^a_{\mu b}(y) }.
\eeq
We integrate out the ghost fields, keeping $A$ fixed, and obtain
\beqa
\label{2propagators}
\vev{ \omega^c_{\nu d}(x) \ \bar\omega^a_{\mu b}(y) }_A & = & - (M^{-1})^{ca}(x, y; A) \delta_{\mu \nu} \delta^{db}\nonumber  \\
\vev{ c^d(x) \ \bar\omega^a_{\mu b}(y)  }_A & = & 0,
\eeqa
where $M^{-1}$ is the inverse of the Faddeev-Popov operator, with kernel $(M^{-1})^{bc}(x, y; A)$.  This gives
\beq
\label{sLiscacul1}
\vev{s \bar\omega^a_{\mu b}(y) } = - \gamma^{1/2}  \frac{L}{2}\sum_{\mu =1}^d\sum_{\sigma=\pm} \int_{x_\mu=\sigma L/2}\hspace{-2em} dS_\mu \  \vev{ (D_\mu M^{-1})^{ba}(x, y; A)  }\ .
\eeq
By symmetries of the periodic hypercube we have
\beq
\vev{ (D_\mu M^{-1})^{ba}(x, y; A) } = {1 \over L^d} {\sum_p} \  f_\mu(p) \ e^{-ip \cdot(x-y)}  \delta^{ba},
\eeq
where $p_\mu = 2 \pi n_\mu / L$, and the $n_\mu$ are integers and where $f_\mu(p)\to p_\mu f(p^2)$ in the infinite-volume limit. By definition, the ghost propagator satisfies,
\beq
\p_\mu (D_\mu M^{-1})^{ba}(x, y; A) = - \delta_0^d(x - y) \delta^{ba},
\eeq
where 
\beq
\delta_0^d(x - y) = { 1 \over L^d} {\sum_p}'  \ e^{-ip \cdot(x-y)} = \delta^d(x - y) - { 1 \over L^d }.
\eeq
The prime on the summation means that the term $p = 0$ is omitted because the constant modes are the trivial null-space of $\p_\mu$.  This gives 
\beq
\vev{ (D_\mu M^{-1})^{ba}(x, y; A)  } =  { 1 \over L^d} \left( {\sum_p}' \ {(- i p_\mu) \over p^2 } \ e^{-ip \cdot(x-y)} + a_\mu \right) \delta^{ba},
\eeq
where $a_\mu$ is an arbitrary constant that we may set to 0.  We thus find,
\beq
\vev{ s \bar\omega^a_{\mu b}(y) }= - \gamma^{1/2} \delta^a_b  \frac{L}{2}\sum_{\mu =1}^d\sum_{\sigma=\pm} \int_{x_\mu=\sigma L/2}\hspace{-2em} dS_\mu \   {1 \over L^d} \left( {\sum_p}' { (- i p_\mu) \over p^2} \ e^{-ip \cdot(x-y)} \right) .
\eeq 
Integrating over the $d - 1$-dimensional  surfaces at $x_\mu=\pm L/2$ sets $p_\perp = 0$ and gives, 
\beq\label{fsum}
\vev{ s \bar\omega^a_{\mu b}(y) } = \gamma^{1/2}  \delta^a_b  \left( {\sum_{p_\mu}}' {  i \over p_\mu} \ e^{-ip_\mu (L/2-y_\mu)} \right)\ ,
\eeq 
where we used that $e^{-ip_\mu(L/2-y_\mu)}=e^{-ip_\mu(-L/2-y_\mu)}$.  The sum over modes in \eqref{fsum} is readily verified to be the fourier representation of the periodic saw-tooth function, 
\beq
\label{gisdefined}
h(y_\mu) \equiv \sum_{n \neq 0}(-1)^n {iL  \over 2 \pi n} \ e^{2 \pi i n y_\mu/L}=y_\mu \ \ {\rm for}\ \ |y_\mu|<L/2.
\eeq
We thus have verified that,
\beq
\vev{ s \bar\omega^a_{\mu b}(y) } = \gamma^{1/2} \delta^a_b y_\mu \ {\rm for}\ \ |y_\mu|<L/2,
\eeq
which in the limit $L \to \infty$ agrees with \eqref{sbomega} for all finite $|y|$.

\section{The criterion implies $\vev{ s D_\mu \bar\omega_{\nu b}} = 0$}
\label{applycriterion}

As an example that will be useful in the accompanying  article~\cite{II}, we apply the criterion~(\ref{sufficient}) to the operator $s D_\mu \bar\omega^a_{\nu b}$.  We need to estimate the asymptotic behavior of 
\beq
C^a_{b\lambda \mu \nu}(x-y) = \vev{\Tr (D_\lambda \omega_\lambda +(D_\lambda c) \times \varphi_\lambda )(x) \ (D_\mu \bar\omega_{\nu b})^a(y)} \hspace{1cm}  {\rm (no \ sum \ on} \ \lambda ),
\eeq
for large $|x|$.   Assuming that the phantom symmetry generated by $Q_{R,\mu a}$ defined in \eqref{QR} remains unbroken, the term  in $c$ does not contribute by \eqref{comegabarprop}, and so
\beq\label{Dw}
C^a_{b\lambda \mu \nu}(x-y) = \vev{\Tr(D_\lambda \omega_\lambda)(x) \ (D_\mu \bar\omega_{\nu b})^a(y)}\ .
\eeq
The same symmetry further implies that, 
\beqa\label{RsymApp}
0 & = & \vev{[ Q_{R,\nu b}\, ,\Tr (D_\lambda \omega_\lambda)(x) (D_\mu \bar c)^a(y) \ ]}
\nonumber \\
& = &  \delta_{\lambda \nu} \vev{ (D_\lambda c)^b(x) (D_\mu \bar c)^a(y) } + i \vev{ \Tr (D_\lambda \omega_\lambda)(x) (D_\mu \bar\omega_{\nu b})^a(y) }\ .
\eeqa
This symmetry therefore implies that, 
\beq
\label{PFaux}
C^a_{b\lambda \mu \nu}(x-y) = - i \delta_{\nu \lambda} \vev{ (D_\lambda c)^b(x) \ (D_\mu \bar c)^a(y) }\ . 
\eeq
The large $|x|$ behavior of the correlator in \eqref{PFaux} is constrained by the horizon condition~(\ref{horizonconditiona}), which reads 
\beq\label{intcrit}
\sum_{a\mu}\int d^dx \ C^a_{a\lambda\mu\mu}(x-y) = (N^2 -1) .
\eeq
The horizon condition thus controls the asymptotic behavior of the correlator $C^a_{b\lambda\mu\nu}(x)$ of \eqref{Dw}, and implies that the criterion of (\ref{sufficient}) is satisfied. The integral of \eqref{intcrit} generally does not converge if the  criterion of \eqref{sufficient} is not fulfilled.    Assuming for instance  that this correlation function  has a power behavior at large $x$,
\beq
\label{asymptoticbehavior}
C^a_{b\lambda\mu\nu}(x) = O( 1 / |x|^p),
\eeq
we obtain $ p > d $ for the integral to converge.  Thus, provided  the horizon condition holds, we conclude that
\beq\label{resA}
\vev{ s(D_\mu\bar\omega_{\nu b})^a } = 0\ .
\eeq
Note that $s(D_\mu\bar\omega_{\nu b})^a$ is not in the space ${\cal W}_{\rm phys}$ of observables defined in \eqref{physicalopsa}, because it is not invariant under the symmetry generated by $Q_{M, \mu\nu}$ of \eqref{QM} and several other phantom symmetries. The class of $s$-exact operators whose expectation value vanishes thus is not limited to the invariant ones in~${\cal W}_{\rm phys}$.

\section{Alternative expression for $ \vev{s F } $}\label{alternativemeansPi}

With~\eqref{hprime},~\eqref{sLbreak} may be written,
\beq
\label{ssubLbreaka}
\vev{s F(y) } = \gamma^{1/2} L \int d^dx \  \sum_{\mu = 1}^d [ 1 - \p_\mu h(x_\mu)] \vev{F(y)\ \Tr (D_\mu \omega_\mu + D_\mu c \times \varphi_\mu) (x) }\ ,
\eeq
and from \eqref{SGZ} we also have that,
\beq
\delta^a_b \ {\delta S \over \delta \bar\omega^a_{\nu b}(x)} = \p_\mu \Tr (D_\mu \omega_\nu + D_\mu c \times \varphi_\nu)(x),
\eeq
which by partial integration gives, 
\beqa
\label{ssubLbreakb}
\vev{s F(y) } & = & \gamma^{1/2} \int d^dx \  \vev{ F(y) \ \Tr\left( s D_\mu \varphi_\mu(x) + h(x_\mu) {\delta S \over \delta \bar\omega_\mu(x) }\right)} 
\nonumber \\
& = & \gamma^{1/2} \int d^dx \  \left[\vev{F(y) \   s \Tr D_\mu \varphi_\mu(x)}+ h(x_\mu) \vev{{\delta F(y) \over \delta \bar\omega^a_{\mu a}(x) }}\right]\ .
\eeqa
The sawtooth function of~\eqref{sawtooth} has the fourier expansion
\beq
h(x_\mu) = i \sum_{n \neq 0} (-1)^n e^{i p_n  x_\mu}/p_n.
\eeq
with $p_n = 2 \pi n/L$, and $n$ is an integer. The first term in
\eqref{ssubLbreakb} thus is the contribution of the zero mode whereas the second contains contributions from nonzero modes only.  In deriving this expression we  used that  non-zero modes may be written as the derivative of a periodic function, whereas the constant zero mode cannot.

\section{Alternative proof that $\langle T_{\mu \nu} \rangle = \langle T_{\mu \nu}^{\rm YM} \rangle$ } \label{testBRST}

We shall show that the BRST-exact part of the energy-momentum tensor has a vanishing expectation value,
\beq
\langle s \Xi_{\mu \nu} \rangle = 0.
\eeq
The proof holds on an infinite or finite periodic hypercubic space-time.
By Lorentz invariance we have $\langle s \Xi_{\mu \nu} \rangle = \delta_{\mu \nu} \langle s \Xi_{\lambda \lambda} \rangle /d)$, where $\langle s \Xi_{\mu \mu} \rangle = (2 - d) \langle s \Psi \rangle$.

It thus is sufficient to show that $\vev{s\Psi}=0$,  where $s\Psi$ is given in~(\ref{SGZ}).  

We use the horizon condition (\ref{eqforgamma}) in~\eqref{SGZ}, and obtain for the $s$-exact part of the trace 
\beqa
\label{Aprime1}
\vev{s\Psi}& = & \Big\langle i \p_\mu b\cdot A_\mu  - i \partial_\mu \bar c\cdot (D_\mu c)
+ \p_\mu \bar\varphi_\nu\cdot (D_\mu \varphi_\nu) 
\\   \nonumber 
&& \ \ \ \ \ \ \ \ \ \   - \p_\mu \bar\omega_\nu\cdot [D_\mu \omega_\nu +  D_\mu c \times \varphi_\nu]
 + \half \gamma^{1/2}\Tr [ D_\mu ( \varphi - \bar\varphi)_\mu - (D_\mu c) \times \bar\omega_\mu]  \Big\rangle, 
\eeqa
where we have divided by the Euclidean volume.  We may integrate by parts without a boundary contribution because of translation invariance and obtain
\beqa
\label{Aprime3}
\vev{s\Psi} & = & \Big\langle  - i b\cdot \p_\mu A_\mu  + i  \bar c\cdot \partial_\mu D_\mu c
- \half \bar\varphi_\nu \cdot (\p_\mu D_\mu \varphi_\nu) +\half\gamma^{1/2} \Tr\, \bar\varphi_\nu \times A_\nu 
\\   \nonumber 
&& \ \ \ \ \ \ \ \ \ \ \ \ \ \ \ \ \ \ 
- \half \varphi_\nu\cdot [D_\mu \p_\mu \bar\varphi_\nu + D_\mu c \times \p_\mu \bar\omega_\nu]-\half \gamma^{1/2}\Tr \,\varphi_\nu \times A_\nu  
\\ \nonumber
&& \ \ \ \ \ \ \ \ \ \ \ 
   + \half \bar\omega_\nu\cdot\p_\mu [D_\mu \omega_\nu + D_\mu c \times \varphi_\nu] -\half \gamma^{1/2}\Tr\,\bar\omega_\nu \times D_\nu c
   -\half\omega_\nu \cdot (D_\mu \p_\mu \bar\omega_\nu)  \Big\rangle. 
\eeqa 
We now use the equations of motion to write this as
\beq
\label{Aprime4}
\vev{s\Psi} =-\int d\Phi  \left(  b\cdot {\delta \over \delta b}  +   \bar c\cdot  {\delta \over \delta \bar c^a}
+ \half  \bar\varphi_\nu\cdot{\delta \over \delta \bar\varphi_\nu}
+ \half  \varphi_\nu\cdot {\delta \over \delta \varphi_\nu}
   + \half \bar\omega_\nu\cdot {\delta \over \delta \bar\omega_\nu}  
   +  \half \omega_\nu\cdot{\delta \over \delta \omega_\nu}  \right) \exp(-S). 
\eeq 
We now do a functional integral by parts and obtain
\beq
\vev{s\Psi} = \delta^d(0) [ (N^2 -1)( 1 - 1) + d(N^2-1)^2( \half + \half - \half - \half) ] = 0,
\eeq
which vanishes because of the opposite statistics of fermions and bosons.  This result can also be verified at tree level.

\end{document}